\documentclass[journal]{IEEEtran}

\usepackage{url}
\usepackage{graphicx}
\usepackage{amsmath}
\usepackage{amssymb}
\usepackage{amsmath,amsthm}
\newtheorem{theorem}{Theorem}
\usepackage{algorithm}
\usepackage{algorithmic}
\usepackage{stfloats}
\usepackage{subfigure}
\usepackage{cite}
\usepackage[justification=centering]{caption}
\usepackage{booktabs}  
\usepackage{epstopdf}
\usepackage[T1]{fontenc}

\ifCLASSINFOpdf

\else

\fi

\begin{document}
\title{Worst-Case Riemannian Optimization with Uncertain Target Steering Vector for Slow-Time Transmit Sequence of Cognitive Radar}

\author{Xinyu~Zhang, Weidong~Jiang, Xiangfeng~Qiu*, Yongxiang Liu
    \thanks{This work is supported by the National Natural Science Foundation of China under Grants 61921001 and 62022091. (Corresponding author: Xiangfeng Qiu.)}
\thanks{Xinyu Zhang is with the College of Electronic Science and Technology, National University of Defense Technology, Changsha, China. e-mail: zhangxinyu90111@163.com.}
\thanks{Weidong Jiang, Xiangfeng Qiu and Yongxiang Liu are also with the College of Electronic Science and Technology, National University of Defense Technology. e-mail: qxf1981993100@163.com; jwd2232@vip.163.com; lyx\underline{~}bible@sina.com}
}

\markboth{Journal of \LaTeX\ Class Files,~Vol.~14, No.~8, August~2015}%
{Shell \MakeLowercase{\textit{et al.}}: Bare Demo of IEEEtran.cls for IEEE Journals}
%



\maketitle
\begin{abstract}
Optimization of slow-time transmit sequence endows cognitive radar with the ability to suppress strong clutter in the range-Doppler domain.  However, in practice, inaccurate target velocity information or random phase error would induce uncertainty about the actual target steering vector, which would in turn severely deteriorate the the performance of the slow-time matched filter.  In order to solve this problem, we propose a new optimization method for slow-time transmit sequence design. The proposed method transforms the original non-convex optimization with an uncertain target steering vector into a two-step worst-case optimization problem.  For each sub-problem,  we develop a corresponding trust-region Riemannian optimization algorithm.  By iteratively solving the two sub-problems, a sub-optimal solution can be reached without accurate information about the target steering vector. Furthermore, the convergence property of the proposed algorithms has been analyzed and detailed proof of the convergence is given.  Unlike the traditional waveform optimization method, the proposed method is designed to work with an uncertain target steering vector and therefore, is more robust in practical radar systems. Numerical simulation results in different scenarios verify the effectiveness of the proposed method in suppressing the clutter and show its advantages in terms of the output signal-to-clutter plus noise ratio (SCNR) over traditional methods.
\end{abstract}


\IEEEpeerreviewmaketitle

\section{Introduction}

\IEEEPARstart{W}{aveform} optimization for active arrays is a rapidly developing field in cognitive radar (CR) with widespread applications. Unlike traditional radar systems, CR systems intelligently interact with the environment and the targets by leveraging historical observations and prior information datasets. This allows for adaptive adjustment of transmit waveforms or their parameters based on a predefined optimization criterion\cite{1593335}. The optimization of transmit waveforms has been extensively explored by researchers, as it constitutes a crucial component of CR systems. Broadly speaking, existing studies on transmit waveform design can be categorized into the following areas.

The first category involves designing a transmit waveform with a small integrated sidelobe level (ISL). In practical radar systems, it is crucial to minimize the energy of the auto-correlation sidelobes to avoid the adverse effects of strong clutter sources \cite{10089242,9628070,9329145}. 
Recent research has focused on developing waveforms with low ISL for both single-channel and multiple-channel radars. Various optimization frameworks, such as majorization-minimization (MM) \cite{7605511,7420715, 9455330}, coordinate descent (CD) \cite{7967829,8835518,9337317,8815508}, gradient descent (GD) \cite{8168352,7485309,9684877}, alternating direction method of multipliers (ADMM) \cite{7529179, 8764367, 9431655}, and inexact alternating direction penalty method (IADPM) \cite{BU2021108145}, have been employed to minimize the ISL. These frameworks aim to improve the target detection performance of radar systems by minimizing the auto-correlation sidelobes. Additionally, another approach to enhance target detection performance is by designing waveforms with low peak sidelobe level (PSL), which refers to the maximum sidelobe of the transmit waveform. In \cite{7967829}, the joint minimization of ISL and PSL is achieved using the CD framework under the constraints of continuous and discrete phases. The proposed algorithm initializes by considering the $l_p$-norm of auto-correlation sidelobes with increasing values of $p$. Similarly, MM-based approaches \cite{7362231} have also been proposed for designing transmit waveforms with small sidelobes based on $l_p$-norm criteria for single-channel radar systems. These results demonstrate that by gradually increasing $p$ during the minimization process, transmit waveforms with very small PSL values can be obtained.
Summarizing the aforementioned works, it is necessary to have an ideal target steering vector and a precise distribution of clutter in the range dimension to establish the waveform design model.

In the pulse-Doppler radar system, strong clutter can obscure the real targets not only in the range dimension but also in the Doppler dimension \cite{9714285}. The ambiguity function (AF) is a widely used two-dimensional performance metric for radar waveforms \cite{doi:https://doi.org/10.1002/0471663085.ch3}. Ideally, a perfect AF should have a "thumbtack" shape, characterized by a relatively high peak at the origin and a wide, shallow base \cite{He_Li_Stoica_2012}. However, achieving this ideal AF is challenging due to the properties of constant energy and constant volume \cite{8961364}. As a result, researchers have turned to designing waveforms with specific shapes that align with the clutter distribution in range and Doppler dimensions.

The research on waveform design to shape the corresponding AF can generally be classified into two categories based on the code nature. The first category focuses on shaping the fast-time AF (FTAF), which aims at designing transmit waveforms composed of a series of sub-pulses. In work \cite{1057703}, the waveform design is achieved by minimizing the integrated square error (ISE) between the realized FTAF and the desired FTAF. The algorithm is known for its computational efficiency and availability of analytical solutions. However, the desired FTAF should be rigorously designed according to the clutter properties.
In literature \cite{7485309}, an algorithm named accelerated iterative sequential optimization (AISO) is proposed to locally shape the FTAF over specific range-Doppler areas. By designing transmit waveforms with low-response values in the area of interest, the probability of detection (PoD) for targets can be further improved. This algorithm actually falls within the MM framework. 
In addition, some studies have extended the FTAF shaping problem to radar systems with multiple channels. In \cite{7738627}, the local auto-AFs and cross-AFs of a set of transmit waveforms are optimized using an algorithm that combines the efficient energy descent method with a fast Fourier transform (FFT). Additionally, a multistage AISO (MS-AISO) algorithm \cite{7738627} is proposed to design orthogonal transmit waveforms with smaller sidelobe levels compared to the method described in \cite{8746293}. 
To overcome the strict constant modulus (CM) constraints imposed on the transmit waveform, the authors in literature \cite{9684877} employ the GD strategy to design the waveform set under low peak-to-average ratio (PAR) constraints \cite{8720029,9855403}. In these previously mentioned works, to obtain a desirable waveform optimization performance, the targets are all assumed to be ideal point targets, and the Doppler steering vectors are assumed to be precisely known.

The second category focuses on slow-time waveform design, which is related to the range-Doppler response and can be interpreted as the slow-time ambiguity function (STAF). In this context, "slow-time" refers to the transmit radar waveform that consists of a series of pulses with a constant pulse repetition interval (PRI). The STAF can be seen as different "slices" of the fast-time counterpart, with a time delay of zero and multiples of a specific number of PRI.
Similar to the waveform design problem in the fast-time dimension, the STAF shaping problem aims to minimize clutter energy in specific range-Doppler bins. In literature \cite{6563125}, the author discusses the STAF shaping problem and formulates it as a non-convex quartic program that is related to the slow-time waveform. To find the optimal solution, an optimization procedure based on the maximum block improvement (MBI) strategy \cite{doi:10.1137/130939110} is proposed.
Another approach to solving the STAF shaping problem is through the combination of the MM framework and CD strategy, as proposed in \cite{7707372}. The waveform design is first modeled as a fractional program, then it is transformed into a polynomial optimization problem under the assumption of an ideal target steering vector. 
This novel algorithm helps overcome the trade-off between shaping performance and computational complexity.
To address the challenges caused by non-convex constraints, researchers have introduced Riemannian manifold-based algorithms into the radar waveform design field. 
In literature \cite{8793214}, the authors transform the CM-constraint optimization problem into an unconstrained one in Riemannian space. A gradient descent algorithm based on the Riemannian first-order gradient is then derived to directly address the resultant problem. Additionally, different second-order optimization algorithms are further studied in \cite{9714285,10164148} to improve the optimization efficiency.
Furthermore, some works extend the AF shaping problem to radar systems with multiple channels \cite{7485309, 9684877, 8746293}. 
To establish the waveform design model for STAF shaping, the previously involved assumptions of ideal point targets and steering vectors are still needed.

In the aforementioned works on waveform design, the target of interest is usually approximated as an ideal point target with a precise steering vector. However, for practical applications, due to the estimation error of target velocity and random phase errors of radar system, the target steering vector may not strictly conform to the ideal assumption. Then, there will inevitably be a significant degradation in the application performance of the designed waveform.
Thus, the designed waveform is required to exhibit robust performances against the inaccurate prior information of the target. In literature \cite{6404093}, in order to improve the target detection performance under the condition of imprecise target Doppler steering vector, the author established a joint design model of the transmission and receiving by maximizing the worst-case output SINR. To solve challenges caused by the considered non-convex constraints, they further proposed a sequential optimization strategy to effectively search for the optimal pair of waveform and filter. Due to the inevitable computation of semidefinite programming (SDP), the proposed algorithm suffers from a high computation burden. Considering the uncertainties of the target parameters, the author \cite{ZHOU2020102709} built a robust waveform design model by maximizing the energy-constrained average SINR. There are also some other robust works on space-time adaptive processing (STAP) in terms of target uncertainty and clutter uncertainty in \cite{2013Robust, 2018Robust, 2019Robust, 2022Robust}. However, all these methods do not take the continuity of the target uncertainty set into consideration and do not tackle the original non-convex problem directly but seek the sub-optimal solution through solving the relaxed convex counterpart. 

In this article, we focus on addressing the issue of robust waveform design in CR systems with an inaccurate target steering vector in the slow-time dimension. Unlike the traditional methods, we consider the uncertainty about the target steering vector as a continuous set and seek the optimal solution using the Riemannian optimization methods. Compared with the state of the art, the proposed method fits the actual situation in radar systems and provide a different perspective in directly solving non-convex optimization problems with uncertainty constraint. The main contributions of the article are summarized as follows 

\textbf{(1)} \emph{\textbf{Proposing a worst-case paradigm for waveform optimization with an uncertain target steering vector:}}
To improve the clutter suppression performance of cognitive radar, we propose a novel slow-time transmit sequence optimization approach for the case of an uncertain target steering vector. In contrast to traditional waveform design methods, our approach incorporates a bounded ball constraint to account for the distortion of the target steering vector. It then transforms the original non-convex optimization with an uncertain set of target steering vectors into a two-step worst-case paradigm which is particularly effective when the target steering vector information is inaccurate or distorted by random phase error.

\textbf{(2)} \emph{\textbf{Developing Riemannian Trust Region (RTR) Algorithms:}}
The proposed worst-case paradigm divides the original optimization into two sub-problems. To effectively solve the established sub-problems, we leverage the geometric properties of the CM constraints and develop two RTR algorithms respectively. These algorithms transform the original constrained optimization into unconstrained ones in manifold space and search for the minimizer with the derived Riemannian gradient and Hessian matrix. 

\textbf{(3)} \emph{\textbf{Convergence analysis:}}
We conduct a theoretical analysis of the monotonically decreasing property of the proposed algorithms. Specifically, we prove the global convergence of Algorithm 1 and the local convergence of Algorithm 2. The overall convergence property of the proposed worst-case method is analyzed in numerical simulation.

\textit{Notation:} $\mathbb{R}^{n}$ and $\mathbb{C}^{n}$ denote the $n$-dimensional real and complex vector space, respectively. $\mathbb{R}^{n\times n}$ and $\mathbb{C}^{n\times n}$ denote the $n\times n$ real and complex matrix space, respectively. Boldface upper case letters stand for matrices while boldface lower case letters stand for column vectors. Standard lowercase letters stand for scalars. $\odot$ denotes the Hadamard product. $\mathbb{E}(\cdot)$ denotes the statistical expectation. $\mathbf{x}^{T}$, $\mathbf{x}^{\ast}$, $\mathbf{x}^{H}$ denote the transpose, complex conjugate, and conjugate transpose respectively. $\text{diag}(\cdot)$ is an operator that transforms a vector into its corresponding diagonal matrix. $|\cdot|$ denotes the modulus of a complex. $\|\cdot\|$ denotes the $l_{2}$-norm of a vector. $\mathbf{1}$ is a column vector whose elements are all 1. $\mathbf{I}_{N\times N}$ represents the $N\times N$ identity matrix. $\lambda_{min}(\cdot) $means the minimum eigenvalue of a matrix, and $min \left( a, b \right)$ outputs the smaller one between $a$ and $b$. $Re\left(\cdot \right)$ and $Im \left(\cdot \right)$ denote the real part and the image part of a scalar, vector, or matrix respectively. $d \left( \cdot \right)$ is the derivation operator. $j$ is an imaginary unit with $j=\sqrt{-1}$.

\section{Signal Model and Problem Formulation}
\label{section: II}
\subsection{Signal Model}
Consider a monostatic pulse-Doppler radar that transmits a burst of $N$ coherent pulses. Each pulse adopts the same waveform except that the initial phases of the pulse train are modulated by a designed phase sequence. Let us denote such sequence by $\mathbf{s} = [s(1),\cdots,s(N)]^{T}\in \mathbb{C}^{N}$. The received pulses first undergo linear matched filtering pulses, then they are processed by a slow time matched filtering that suppresses unwanted clutter with Doppler frequencies different from the target.

Assuming the target normalized Doppler frequency is $v_{t}$, then the Doppler steering vector of the target can be written as $\mathbf{p}(v_{t})=\left[ 1, e^{j2\pi v_{t}}, \cdots, e^{j2\pi (N-1)v_{t}}\right]^{T}$. Denote $\alpha_{t}$ as a complex parameter that accounts for channel propagation and backscattering effects from the target. The received signal before the slow-time matched filtering can be modeled as
\begin{equation}
\label{equ1}
\mathbf{y} = \alpha_{t}\mathbf{s}\odot\mathbf{p}(v_{t}) + \mathbf{c}+\mathbf{n},
\end{equation}
where $\mathbf{n}$ represents the system noise and $\mathbf{c}$ is the signal-dependent clutter from the environment.

The clutter vector $\mathbf{c}$ is the superposition of echoes from different uncorrelated scatterers distributed in different range-azimuth bins $(r,l)$ with $r \in \left\{ 0,1,\cdots,N_{c}-1\right\}$, $l\in \left\{0,1,\cdots,L-1\right\}$ where $N_{c}\leq N$ is the number of range bins and $L$ is the number of azimuth sectors. From the perspective of radar signal processing, each range-azimuth bin corresponds to a range-Doppler bin, thus returns from the $(r,l)$ range-azimuth sector have the Doppler steering vector $\mathbf{p}(v_{c_{k}})$ where $v_{c_{k}}$ is the normalized Doppler frequency of the $k\text{-th}$ scattering points. There is a one-to-one map from $(r,l)\in \left\{0,1,\cdots,N_{c}-1\right\}\times \left\{0,1,\cdots,L-1\right\}$ to $k \in \left\{0,1,\cdots,N_{c}L-1\right\}$. Considering the range ambiguity, scattering points in adjacent range positions may interfere with the target echo. As a consequence, the clutter can be expressed as
\begin{equation}
\label{equ2}
\mathbf{c} = \sum_{k=0}^{N_{t}-1}\varrho_{k}\mathbf{J}^{r_{k}}\left(\mathbf{s}\odot\mathbf{p}(v_{c_{k}})\right),
\end{equation}
where $N_{t}\leq N_{c}L$ is the total number of interfering scatterers, $r_{k} \in \left\{0,1,\cdots,N-1\right\}$ is the range ambiguity number, $\varrho_{k}$ accounts for the complex amplitude of the $k\text{-th}$ scatterer, and $\mathbf{J}^{r_{k}}$ denotes the shift matrix, i.e. $\forall r \in \left\{0,1,\cdots,N-1\right\}$
\begin{equation}
\label{equ3}
\mathbf{J}^{r_{k}}(m,n)=\left\{\begin{matrix} 1, & m-n=r \\ 0, & m-n \neq r \end{matrix}\right. .
\end{equation}

The output of the slow-time matched filter to the target signal $\mathbf{s}\odot \mathbf{p}(v_{t})$ can be expressed as
\begin{equation}
\label{equ4}
\begin{split}
(\mathbf{s}\odot \mathbf{p}(v_{t}))^{H}\mathbf{y} &= \alpha_{t}\left\|\mathbf{s}\right\|^{2} +  (\mathbf{s}\odot \mathbf{p}(v_{t}))^{H}\mathbf{n} \\
&+\sum_{k=0}^{N_{t}-1}\varrho_{k}(\mathbf{s}\odot \mathbf{p}(v_{t}))^{H}\mathbf{J}^{r_{k}}\left(\mathbf{s}\odot\mathbf{p}(v_{c_{k}})\right)
\end{split}.
\end{equation}

It is commonly assumed that the noise $\mathbf{n}$, uncorrelated from $\mathbf{c}$, is zero-mean complex Gaussian white noise with covariance matrix $\mathbb{E}(\mathbf{n}\mathbf{n}^{H}) = \sigma_{n}^{2}\mathbf{I}$. Consequently, the disturbance power at the output of the matched filter can be expressed as
\begin{equation}
\label{equ5}
\begin{split}
\mathbb{E}&[\lvert(\mathbf{s}\odot \mathbf{p}(v_{t}))^{H}\mathbf{n}+\sum_{k=0}^{N_{t}-1}\varrho_{k}(\mathbf{s}\odot \mathbf{p}(v_{t}))^{H}\mathbf{J}^{r_{k}}\left(\mathbf{s}\odot\mathbf{p}(v_{c_{k}})\right)\rvert^{2}]\\
&= \sigma_{n}^{2}\| \mathbf{s}\|^{2} +\sum_{k=0}^{N_{t}-1}\lvert\mathbf{s}^{H}\mathbf{\Psi}_{k}\mathbf{s}\rvert^{2}
\end{split}
\end{equation}
where  $\mathbf{\Psi}_{k}=\sqrt{\sigma_{k}^{2}}\text{diag}(\mathbf{p}(v_{t}))\mathbf{J}^{r_{k}}\text{diag}(\mathbf{p}(v_{c_{k}}))$ and $\sigma_{k}^{2}=\mathbb{E}(\varrho_{k}\varrho_{k}^{*})$. Then the signal to clutter plus noise ratio (SCNR) can be calculated as
\begin{equation}
\label{equ6}
\text{SCNR}=\frac{\sigma_{t}^{2}\|\mathbf{s}^{H}\mathbf{s}\|^{2}}{\sigma_{n}^{2}\| \mathbf{s}\|^{2} +\sum_{k=0}^{N_{t}-1}\lvert\mathbf{s}^{H}\mathbf{\Psi}_{k}\mathbf{s}\rvert^{2}}
\end{equation}
with $\sigma_{t}^{2}=\mathbb{E}(\alpha_{t}\alpha_{t}^{*})$.

\subsection{Problem Formulation}
In the CR system, the slow-time transmit sequence can be optimized given the prior information about the clutter distribution. Based on (\ref{equ6}), to characterize the SCNR, we need the mean power $\sigma_{k}^{2}$ as well as the Doppler frequencies $v_{t}$ and $v_{c_{k}}$ of each scatterer. Such information can be obtained by the cognitive radar system in advance. The mean power of the clutter scatterer can be predicted through the link between a digital terrain map and the Radar Cross Section (RCS) model of clutter \cite{8961364, 8398580}. Given the velocity of radar, the Dopper frequency of each clutter scatterer can be estimated from the geometrical relation between the radar and the scatterer.

In the following, without loss of generality, we center the Doppler frequency axis to the target Doppler frequency, namely all the normalized Doppler frequencies are expressed in terms of the difference to $v_{t}$. In such case, $\mathbf{p}(v_{t}) = \mathbf{1}$ where $\mathbf{1}$ is a vector whose elements are all $1$. However, in practice, the actual Doppler steering vector of the target is not always equal to our presumed $v_{t}$. Such difference could be induced by the estimation error of target velocity or by the random phase noise of the radar system. Denote the actual Doppler steering vector of the target by $\tilde{\mathbf{p}}(v_{t})$ and $\tilde{\mathbf{s}} = \mathbf{s}\odot \tilde{\mathbf{p}}(v_{t})$. Since the phase error does not impact the amplitude, we have $\lvert\tilde{\mathbf{s}}(i)\rvert= 1,\forall i \in \mathcal{N}=[1,2,\cdots,N]$.

Considering the above phase error, we formulate our problem as
\begin{equation}
\label{equ7}
\begin{split}
\underset{\mathbf{s}}{\mathop{\max }}\,&\frac{\lvert\mathbf{s}^{H}\tilde{\mathbf{s}}\rvert^{2}}{\sigma_{n}^{2}\| \mathbf{s}\|^{2} +\sum_{k=0}^{N_{t}-1}\lvert\mathbf{s}^{H}\mathbf{\Psi}_{k}\mathbf{s}\rvert^{2}},\\ &\text{s.t.} \,\lvert\mathbf{s}(i)\rvert=1, \forall i \in \mathcal{N}, \tilde{\mathbf{s}}\in\mathcal{A}
\end{split}
\end{equation}
where $\mathcal{A}$ is a set of possible target steering vectors. Assume that the distorted target  steering vector is bounded in a ball around the presumed target steering vector, thus $\mathcal{A}$ can be expressed as
\begin{equation}
\label{equ8}
\mathcal{A}= \left\{\tilde{\mathbf{s}}|\|\tilde{\mathbf{s}}-\mathbf{s}\|^{2}\le \varepsilon, \lvert\tilde{\mathbf{s}}(i)\rvert=1, \forall i \in \mathcal{N} \right\}.
\end{equation}
In equation (\ref{equ7}), the unit norm constraint on the transmit sequence is imposed due to the operating conditions of radar amplifiers. Radar amplifiers typically work in saturation conditions, which prohibit the use of amplitude modulation in radar waveforms.

Because the distorted target steering vector is unknown, we adopt the worst-case paradigm to solve the problem. Thus, the problem (\ref{equ7}) can be decomposed into a two-step optimization, i.e.
\begin{subequations}
\label{equ9}
\begin{equation}
\label{equ9.1}
\underset{\tilde{\mathbf{s}}}{\mathop{\min }}\,\lvert\mathbf{s}^{H}\tilde{\mathbf{s}}\rvert^{2},\quad \text{s.t.} \quad\|\tilde{\mathbf{s}}-\mathbf{s}\|^{2}\le \varepsilon, \lvert\tilde{\mathbf{s}}(i)\rvert=1, \forall i \in \mathcal{N}
\end{equation}
\begin{equation}
\label{equ9.2}
\underset{\mathbf{s}}{\mathop{\max }}\,\frac{\lvert\mathbf{s}^{H}\tilde{\mathbf{s}}\rvert^{2}}{\sigma_{n}^{2}\| \mathbf{s}\|^{2} +\sum_{k=0}^{N_{t}-1}\lvert\mathbf{s}^{H}\mathbf{\Psi}_{k}\mathbf{s}\rvert^{2}}, \text{s.t.} \,\lvert\mathbf{s}(i)\rvert=1, \forall i \in \mathcal{N},
\end{equation}
\end{subequations}
The first step is to find the worst case for the current transmit sequence while the second step is to optimize the transmit sequence for the worst case.

\textit{Remark 1}: Note that the distorted target steering vector $\tilde{\mathbf{s}}$ is assumed to be distributed within a ball centered around the designed transmit sequence $\mathbf{s}$. Therefore, the solution of each sub-problem would impact the solution of the other. If the distribution of phase error is a priori, we could derive the solution to (\ref{equ7}) through the Bayesian approach.

To obtain a final design of the transmit sequence, we would need to iteratively optimize $\mathbf{s}$ and $\tilde{\mathbf{s}}$ until the output SCNR reaches a relatively stable level.

In practice, the noise power $\sigma_{n}^{2}$ is usually much smaller than the clutter power. Therefore, for most radar systems, the more imminent task is to improve the signal-to-clutter ratio (SCR) instead.

\textit{Remark 2}: In fact, the noise term in (\ref{equ9.2}) is a constant, given the constraint $\lvert\mathbf{s}(i)\rvert=1, \forall i \in \mathcal{N}$.

\textit{Remark 3}: It should be noted that our transmit sequence is designed for the slow-time axis in pulse Doppler radar. Before the slow-time matched filtering, the signal-to-noise ratio (SNR) has already been largely increased through the fast-time matched filtering.

Therefore, we could rewrite the two-step optimization problem as
\begin{subequations}
\label{equ10}
\begin{equation}
\label{equ10.1}
\underset{\tilde{\mathbf{s}}}{\mathop{\min }}\,\lvert\mathbf{s}^{H}\tilde{\mathbf{s}}\rvert^{2},\quad \text{s.t.} \quad\|\tilde{\mathbf{s}}-\mathbf{s}\|^{2}\le \varepsilon, \lvert\tilde{\mathbf{s}}(i)\rvert=1, \forall i \in \mathcal{N}
\end{equation}
\begin{equation}
\label{equ10.2}
\underset{\mathbf{s}}{\mathop{\max }}\,\frac{\lvert\mathbf{s}^{H}\tilde{\mathbf{s}}\rvert^{2}}{\sum_{k=0}^{N_{t}-1}\lvert\mathbf{s}^{H}\mathbf{\Psi}_{k}\mathbf{s}\rvert^{2}}, \text{s.t.} \,\lvert\mathbf{s}(i)\rvert=1, \forall i \in \mathcal{N},
\end{equation}
\end{subequations}

\section{Optimizing Transmit Sequence with Random Phase Error}
\label{section: III}
To begin with, it can be found in (\ref{equ10}) that the solution to the formulated problem is confined in a smooth manifold $\mathcal{M}$ composed of complex unit circles. Thus we adopt the Riemannian optimization method to solve (\ref{equ10}).

The differential of all the possible curves on the manifold at a certain point $\mathbf{x} \in \mathcal{M}$ forms a linear space, namely tangent space $T_{\mathbf{x}}\mathcal{M}$. Any tangent vector $\xi_{\mathbf{x}}$ in the tangent space is a mapping
\begin{equation}
\label{equ11}
\mathbf{\xi}_{\mathbf{x}}f = \frac{d(f(\gamma(t)))}{dt}\Big|_{t=0},\forall f \in \chi_{\mathbf{x}}\mathcal{M}
\end{equation}
where $\chi_{\mathbf{x}}\mathcal{M}$ denotes the set of smooth real-valued functions defined on a neighborhood of $\mathbf{x}$. Since we impose the solution on $\mathcal{M}$, curves through the point $\mathbf{x}$ characterize the possible traces of moving the variable.
The tangent space of our manifold can be expressed as
\begin{equation}
\label{equ12}
T_{\mathbf{x}}\mathcal{M}=\left\{\mathbf{\xi}_{\mathbf{x}} \in \mathbb{C}^{N}:\mathop{Re}(\mathbf{\xi}_{\mathbf{x}}\odot\mathbf{x}^{*})=\mathbf{0},\mathbf{x}\in\mathcal{M}\right\}
\end{equation}
The derivation of the tangent space can be found in Appendix \ref{app1}.

To continue, we endow the tangent space with the Riemannian metric
\begin{equation}
\label{equ13}
g_{\mathbf{x}}(\mathbf{\xi}_{\mathbf{x}},\mathbf{\varsigma}_{\mathbf{x}}) = \left\langle \mathbf{\xi}_{\mathbf{x}},\mathbf{\varsigma}_{\mathbf{x}} \right\rangle, \forall \mathbf{\xi}_{\mathbf{x}},\mathbf{\varsigma}_{\mathbf{x}} \in T_{\mathbf{x}}\mathcal{M}
\end{equation}
where $\left\langle \cdot \right\rangle$ is the canonical complex Euclidean inner product. This choice makes the manifold $\mathcal{M}$ a Riemannian submanifold of Euclidean space $\mathbb{C}^{N}$. Therefore, given a smooth scalar field $f$, the direction of the steepest ascent, namely the Riemannian gradient, is the orthogonal projection from the Euclidean gradient of $f$, $\mathbf{Grad}f(\mathbf{x})$ to the tangent space. Thus, the Riemannian gradient of $f$ can be calculated by
\begin{equation}
\begin{aligned}
    \label{equ14}
\mathbf{grad}f(\mathbf{x}) & = \operatorname{Proj_{\boldsymbol{x}}} \left( \mathbf{Grad}f(\mathbf{x}) \right) \\
& = \mathbf{Grad}f(\mathbf{x})-\mathop{Re}\{\mathbf{Grad}f(\mathbf{x})\odot\mathbf{x}^{*}\}\odot\mathbf{x}
\end{aligned}
\end{equation}
One can easily verify with (\ref{equ14}) that the Riemannian gradient lies in the tangent space.
After moving the variable along the direction of the Riemannian gradient, the retraction maps the result on $T_{\mathbf{x}}\mathcal{M}$ back to the original manifold $\mathcal{M}$. For our manifold, the retraction is defined as
\begin{equation}
\label{equ15}
\text{R}_{\mathbf{x}}(\mathbf{\xi}_{\mathbf{x}})=\frac{\mathbf{x}+\mathbf{\xi}_{\mathbf{x}}}{\|\mathbf{x}+\mathbf{\xi}_{\mathbf{x}}\|}
\end{equation}

On one hand, the tangent space $T_{\mathbf{x}}\mathcal{M}$ provides a local vector space approximation of the manifold. On the other hand, the retractions, a mapping between $\mathcal{M}$ and $T_{\mathbf{x}}\mathcal{M}$ gives us an approach to transform an optimization problem on $\mathcal{M}$ into an optimization problem on linear space $T_{\mathbf{x}}\mathcal{M}$.

\subsection{Solution to (\ref{equ10.1})}
In this subsection, we develop a Riemannian optimization algorithm to solve (\ref{equ10.1}). To start with, we assert the following theorem:

$\textit{Theorem}\, \textit{1}$ :
\begin{theorem}
If $\tilde{\mathbf{s}}^{\star}$ is a plausible solution to (\ref{equ10.1}), it would satisfy $\|\tilde{\mathbf{s}}^{\star}-\mathbf{s}\|^{2}=\varepsilon,\lvert\tilde{\mathbf{s}}^{\star}(i)\rvert=1, \forall i \in \mathcal{N}$.
\end{theorem}

The proof of the theorem can be found in Appendix \ref{app2}. $\textit{Theorem}\, \textit{1}$ states that the minimization of $\lvert\mathbf{s}^{H}\tilde{\mathbf{s}}\rvert^{2}$ can only be achieved at the boundary of the ball $\|\tilde{\mathbf{s}}^{\star}-\mathbf{s}\|^{2}\le\varepsilon$. It allows us to equally express the problem as
\begin{equation}
\label{equ16}
\underset{\tilde{\mathbf{s}}}{\mathop{\min }}\,\lvert\mathbf{s}^{H}\tilde{\mathbf{s}}\rvert^{2},\quad \text{s.t.} \quad\|\tilde{\mathbf{s}}-\mathbf{s}\|^{2}= \varepsilon, \lvert\tilde{\mathbf{s}}(i)\rvert=1, \forall i \in \mathcal{N}.
\end{equation}
From the constraints, one can derive the following
\begin{equation}
\label{equ17}
\mathop{Re}(\mathbf{s}^{H}\tilde{\mathbf{s}})=N-\frac{\varepsilon}{2}.
\end{equation}
Substituting (\ref{equ17}) into (\ref{equ16}), we have
\begin{equation}
\label{equ18}
\underset{\tilde{\mathbf{s}}}{\mathop{\min }}\,\mathop{Im}(\mathbf{s}^{H}\tilde{\mathbf{s}})^{2},\quad \text{s.t.} \mathop{Re}(\mathbf{s}^{H}\tilde{\mathbf{s}})=N-\frac{\varepsilon}{2}, \lvert\tilde{\mathbf{s}}(i)\rvert=1, \forall i \in \mathcal{N}.
\end{equation}
When omitting the unit norm constraint $\lvert\tilde{\mathbf{s}}(i)\rvert=1, \forall i \in \mathcal{N}$, the rest of (\ref{equ18}) is a convex optimization problem and more precisely, a linear optimization problem. Thus, we could rewrite the problem in its Lagrange form
\begin{equation}
\label{equ19}
\underset{\tilde{\mathbf{s}}}{\mathop{\min }}\,\mathop{Im}(\mathbf{s}^{H}\tilde{\mathbf{s}})^{2}+\lambda(\mathop{Re}(\mathbf{s}^{H}\tilde{\mathbf{s}})-N+\frac{\varepsilon}{2})^{2},\, \text{s.t.} \lvert\tilde{\mathbf{s}}(i)\rvert=1, \forall i \in \mathcal{N},
\end{equation}
where $\lambda$ is the Lagrange multiplier. Combining (\ref{equ14}) and (\ref{equ15}), one could easily derive a first-order Riemannian optimization algorithm. However, in this paper, we develop the  Riemannian trust region (RTR) optimization method to solve (\ref{equ19}), because such a second-order Riemannian algorithm usually performs better. Specifically, the trust region method iteratively optimizes the step $\mathbf{\xi}_{\mathbf{x}} \in T_{\mathbf{x}}\mathcal{M}$ for each iteration by solving the following sub-problem
\begin{equation}
\label{equ20}
\underset{\mathbf{\xi}_{\mathbf{x}} \in T_{\mathbf{x}}\mathcal{M}}{\mathop{\min }}\,m_{k}(\mathbf{\xi}_{\mathbf{x}}),\quad \text{s.t.} \|\mathbf{\xi}_{\mathbf{x}}\|_{\mathbf{x}}\le \Delta_{k},
\end{equation}
where $k$ represents the iteration number, $\Delta_{k}$ is the radius of the trust region at iteration $k$, $\|\cdot\|_{\mathbf{x}}$ is the $l_{2}$ norm induced by the Riemannian metric (\ref{equ13}),  $m_{k}(\mathbf{\xi}_{\mathbf{x}}) = f_{\tilde{\mathbf{s}}}(\tilde{\mathbf{s}}_{k})+\left\langle \mathbf{grad}f_{\tilde{\mathbf{s}}}(\tilde{\mathbf{s}}_{k}),\mathbf{\xi}_{\mathbf{x}} \right\rangle+\frac{1}{2} \left\langle \mathbf{Hess}f_{\tilde{\mathbf{s}}}(\mathbf{\xi}_{\mathbf{x}}),\mathbf{\xi}_{\mathbf{x}} \right\rangle$ is an approximation of the pullback $f_{\tilde{\mathbf{s}}}(\text{R}_{\mathbf{x}})$ where
\begin{equation}
\label{equ21}
f_{\tilde{\mathbf{s}}}(\tilde{\mathbf{s}}_{k})= \mathop{Im}(\mathbf{s}^{H}\tilde{\mathbf{s}}_{k})^{2}+\lambda(\mathop{Re}(\mathbf{s}^{H}\tilde{\mathbf{s}}_{k})-N+\frac{\varepsilon}{2})^{2},
\end{equation}
and $\mathbf{Hess}f_{\tilde{\mathbf{s}}}$ is the Riemannian Hessian defined as the covariant derivative of the gradient vector filed with respect to the Riemannian connection. At any point of the manifold, the Hessian defines a linear operator from the tangent space to itself
\begin{equation}
\label{equ22}
\mathbf{Hess}f_{\tilde{\mathbf{s}}}(\mathbf{\xi}_{\mathbf{x}}) = \nabla_{\mathbf{\xi}_{\mathbf{x}}}\mathbf{grad}f_{\tilde{\mathbf{s}}}(\tilde{\mathbf{s}}_{k}),
\end{equation}
where $\nabla$ denotes the Riemannian connection along the direction $\mathbf{\xi}_{\mathbf{x}}$. 

The trust region method approximates the pullback $f_{\tilde{\mathbf{s}}}(\text{R}_{\mathbf{x}})$ by a simpler second-order model in the tangent space and optimizes the step at each point with it. Since such a model is only a local approximation of the pullback, it can only be trusted in a ball around the origin in the tangent space.

To continue, we need to derive the closed-form expression of the Riemannian gradient and the Hessian. Since $\mathcal{M}$ is a submanifold of Euclidean space, the Riemannian gradient $\mathbf{grad}f_{\tilde{\mathbf{s}}}$ can be calculated by (\ref{equ14}) where the Euclidean gradient $\mathbf{Grad}f_{\tilde{\mathbf{s}}}$ can be expressed as
\begin{equation}
\label{equ23}
\mathbf{Grad}f_{\tilde{\mathbf{s}}} = 2j\mathop{Im}(\mathbf{s}^{H}\tilde{\mathbf{s}}_{k})\mathbf{s}+2\lambda(\mathop{Re}(\mathbf{s}^{H}\tilde{\mathbf{s}}_{k})-N+\frac{\varepsilon}{2})\mathbf{s}.
\end{equation}

For a Riemannian submanifold of Euclidean space, the Riemannian Hessian can be calculated by \cite{boumal2023introduction}
\begin{equation}
\label{equ24}
\mathbf{Hess}f_{\tilde{\mathbf{s}}}(\mathbf{\xi}_{\mathbf{x}}) = \text{Proj}_{\tilde{\mathbf{s}}_{k}}(D\overline{\mathbf{grad}}f_{\tilde{\mathbf{s}}}(\mathbf{\xi}_{\mathbf{x}})),
\end{equation}
where $\overline{\mathbf{grad}}f_{\tilde{\mathbf{s}}}$ is a smooth extension of $\mathbf{grad}f_{\tilde{\mathbf{s}}}$ to the Euclidean space, $D\overline{\mathbf{grad}}f_{\tilde{\mathbf{s}}}(\mathbf{\xi}_{\mathbf{x}})$ represents the directional derivative along the direction $\mathbf{\xi}_{\mathbf{x}}$, $\text{Proj}_{\tilde{\mathbf{s}}_{k}}$ is the same orthogonal projection operator as the one in (\ref{equ14}) which projects arbitrary vector $\mathbf{\upsilon}$ to the tangent space $T_{\mathbf{x}}\mathcal{M}$ at $\tilde{\mathbf{s}}_{k}$ by
\begin{equation}
\label{equ25}
\text{Proj}_{\tilde{\mathbf{s}}_{k}}(\mathbf{\upsilon}) =\mathbf{\upsilon} - \mathop{Re}\left\{\mathbf{\upsilon}\odot\tilde{\mathbf{s}}_{k}^{*}\right\}\odot \tilde{\mathbf{s}}_{k}.
\end{equation}
The directional derivative $D\overline{\mathbf{grad}}f_{\tilde{\mathbf{s}}}(\mathbf{\xi}_{\mathbf{x}})$ can be calculated by its definition
\begin{equation}
\label{equ26} D\overline{\mathbf{grad}}f_{\tilde{\mathbf{s}}}(\mathbf{\xi}_{\mathbf{x}})= \lim_{t\to 0}\frac{\overline{\mathbf{grad}}f_{\tilde{\mathbf{s}}}(\tilde{\mathbf{s}}_{k}+t\mathbf{\xi}_{\mathbf{x}})-\overline{\mathbf{grad}}f_{\tilde{\mathbf{s}}}(\tilde{\mathbf{s}}_{k})}{t}.
\end{equation}

Substituting (\ref{equ14}) and (\ref{equ26}) into (\ref{equ24}) and leveraging the fact that $\text{Proj}_{\tilde{\mathbf{s}}_{k}}(\tilde{\mathbf{s}}_{k}) = \mathbf{0}$, one could get
\begin{equation}
\label{equ27}
\begin{aligned}
    \mathbf{Hess}f_{\tilde{\mathbf{s}}}(\mathbf{\xi}_{\mathbf{x}}) & = \text{Proj}_{\tilde{\mathbf{s}}_{k}}(D{\mathbf{Grad}}f_{\tilde{\mathbf{s}}}(\mathbf{\xi}_{\mathbf{x}})) \\
    & - \mathop{Re}\left\{\mathbf{Grad}f_{\tilde{\mathbf{s}}}\odot\tilde{\mathbf{s}}_{k}\right\}\odot \mathbf{\xi}_{\mathbf{x}},
\end{aligned}
\end{equation}
where
\begin{equation}
\label{equ28}
D{\mathbf{Grad}}f_{\tilde{\mathbf{s}}}(\mathbf{\xi}_{\mathbf{x}})= 2j\mathop{Im}(\mathbf{s}^{H}\mathbf{\xi}_{\mathbf{x}})\mathbf{s}+2\lambda\mathop{Re}(\mathbf{s}^{H}\mathbf{\xi}_{\mathbf{x}})\mathbf{s}.
\end{equation}

Now we have all the ingredients to calculate $m_{k}$. To solve (\ref{equ20}), we turn to the truncated conjugate gradient (TCG) method which can solve the sub-problem to global optimality efficiently. To implement the TCG method, we need to further introduce the vector transporter that transports a tangent vector from one tangent space to another. This is done by
\begin{equation}
\label{equ29}
T_{\tilde{\mathbf{s}}_{k+1}\gets\tilde{\mathbf{s}}_{k}}(\mathbf{\xi}_{\mathbf{x}}) = \mathbf{\xi}_{\mathbf{x}}-\mathop{Re}\left\{\mathbf{\xi}_{\mathbf{x}}^{*}\odot \tilde{\mathbf{s}}_{k+1}\right\}\odot \tilde{\mathbf{s}}_{k+1}
\end{equation}

After solving (\ref{equ20}), the tentative next iterate is $\tilde{\mathbf{s}}_{k+1}=\textbf{R}_{\tilde{\mathbf{s}}_{k}}(\mathbf{\xi}_{\mathbf{x}}^{\star})$ where $\mathbf{\xi}_{\mathbf{x}}^{\star}$ is the optimized step yielded by TCG.

The next step is to determine whether we accept or reject the candidate $\mathbf{\xi}_{\mathbf{x}}^{\star}$. To do this, we need to determine whether our approximation model deviates far from the pullback. This is done by computing the following ratio
\begin{equation}
\label{equ30}
\rho_{k} = \frac{f_{\tilde{\mathbf{s}}}(\tilde{\mathbf{s}}_{k})-f_{\tilde{\mathbf{s}}}(\tilde{\mathbf{s}}_{k+1})}{m_{k}(\mathbf{0})-m_{k}(\mathbf{\xi}_{\mathbf{x}}^{\star})}.
\end{equation}
If $\rho_{k}$ is exceedingly small, the model is very inaccurate. Then we reject the step and reduce the trust region radius $\Delta_{k}$. If $\rho_{k}$ is small but not dramatically so, the step is accepted but the trust region radius is reduced. If $\rho_{k}$ is close to $1$, the model is good and the trust-region radius can be expanded.

The proposed procedure above can be formalized as the following algorithm. Note that practical stopping criteria for the above algorithm usually involve an upper-bound on the total number of iterations and a threshold on the gradient norm. Typically, the threshold $\epsilon$ can be set to $10^{-8}\|\mathbf{grad}f_{\tilde{\mathbf{s}}}(\tilde{\mathbf{s}}_{0})\|_{\tilde{\mathbf{s}}_{0}}$. The RTR method also has good convergence properties, which we will analyze later.
\begin{figure}[H]
\begin{tabular}{p{8.5cm}}
\hline
\textbf{Algorithm 1: RTR method to solve (\ref{equ10.1})} \\
\hline
\begin{enumerate}
\item \textbf{Parameters}: maximum trust region radius $\bar{\Delta}$, threshold $\bar{\rho}\in(0,\frac{1}{4})$, user-determined threshold $\epsilon$.
  \item \textbf{Input}: $\tilde{\mathbf{s}}_{0}\in \mathcal{M}, \Delta_{0}\in (0,\bar{\Delta}]$
  \item \textbf{For} $k=0,1,2,\cdots$
  \begin{enumerate}
  \item Calculate the Riemannian gradient at $\tilde{\mathbf{s}}_{k}$ using (\ref{equ14}) and (\ref{equ23}), formalize the Hessian map by (\ref{equ23}), (\ref{equ27}) and (\ref{equ28}).
  \item Use TCG method to solve the sub-problem (\ref{equ20}), yielding $\mathbf{\xi}_{\mathbf{x}}^{\star}$.
  \item Compute $\rho_{k}$ by (\ref{equ30}).
  \item Accept or reject the tentative next iterate:
  \begin{equation}
    \label{equ31}
    \tilde{\mathbf{s}}_{k+1}=\left\{\begin{matrix} \textbf{R}_{\tilde{\mathbf{s}}_{k}}(\mathbf{\xi}_{\mathbf{x}}^{\star}) & \text{if}\, \rho_{k}>\bar{\rho}\, \text{(accept)} \\ \tilde{\mathbf{s}}_{k} & \text{otherwise (reject)} \end{matrix}\right.
  \end{equation}
  \item Update the trust-region radius:
   \begin{equation}
    \label{equ32}
    \Delta_{k+1}=\left\{\begin{matrix} \frac{1}{4}\Delta_{k} & \text{if}\, \rho_{k}<\frac{1}{4} \\ \mathop{min}(2\Delta_{k},\bar{\Delta}) & \begin{split} & \text{if}\, \rho_{k}>\frac{3}{4}\, \text{and}\\ &\|\mathbf{\xi}_{\mathbf{x}}^{\star}\|_{\tilde{\mathbf{s}}_{k}}=\Delta_{k} \end{split}  \\ \Delta_{k} & \text{otherwise} \end{matrix}\right.
  \end{equation}
  \item Until $\|\mathbf{grad}f_{\tilde{\mathbf{s}}}\|_{\tilde{\mathbf{s}}_{k+1}} \le \epsilon$.
  \end{enumerate}
  \item \textbf{Output}: output $\tilde{\mathbf{s}}_{k+1}$.
\end{enumerate}\\
\hline
\end{tabular}
\end{figure}

\subsection{Solution to (\ref{equ10.2})}
In this subsection, we develop an RTR method to solve (\ref{equ10.2}). The main procedure is similar to the one derived in solving (\ref{equ10.1}). Note that the optimizing variable $\mathbf{s}$ lies on the same manifold $\mathcal{M}$ as $\tilde{\mathbf{s}}$, thus we adopt the same symbols to denote the tangent space and the tangent vectors for this subsection. The choice of Riemannian metric remains the same as (\ref{equ13}). The main step in solving (\ref{equ10.2}) involves the following sub-problem
\begin{equation}
\label{equ33}
\underset{\mathbf{\xi}_{\mathbf{x}} \in T_{\mathbf{x}}\mathcal{M}}{\mathop{\min }}\,h_{k}(\mathbf{\xi}_{\mathbf{x}}),\quad \text{s.t.} \|\mathbf{\xi}_{\mathbf{x}}\|_{\mathbf{x}}\le \Delta_{k},
\end{equation}
where 
\begin{equation}
\begin{aligned}
\nonumber
    h_{k}(\mathbf{\xi}_{\mathbf{x}}) = f_{\mathbf{s}}(\mathbf{s}_{k}) &+ \left\langle \mathbf{grad}f_{{\mathbf{s}}}({\mathbf{s}}_{k}),\mathbf{\xi}_{\mathbf{x}} \right\rangle \\
    &+\frac{1}{2} \left\langle \mathbf{Hess}f_{{\mathbf{s}}}(\mathbf{\xi}_{\mathbf{x}}),\mathbf{\xi}_{\mathbf{x}} \right\rangle
\end{aligned}
\end{equation} 
is an approximation of the pullback $f_{{\mathbf{s}}}(\text{R}_{\mathbf{x}}(\mathbf{\xi}_{\mathbf{x}}))$ where
\begin{equation}
\label{equ34}
f_{{\mathbf{s}}}({\mathbf{s}}_{k})= \frac{\sum_{i=0}^{N_{t}-1}\lvert\mathbf{s}^{H}_{k}\mathbf{\Psi}_{i}\mathbf{s}_{k}\rvert^{2}}{\lvert\mathbf{s}^{H}_{k}\tilde{\mathbf{s}}\rvert^{2}}
\end{equation}
and $\mathbf{s}_{k}$ is the variable to be optimized at the $k-\text{th}$ iteration. To solve the sub-problem (\ref{equ33}), we need to derive the expression of the Riemannian gradient and Hessian. The Riemannian gradient can be calculated by
\begin{equation}
\label{equ35}
\mathbf{grad}f_{\mathbf{s}}= \text{Proj}_{\mathbf{s}}(\mathbf{Grad}f_{\mathbf{s}}),
\end{equation}
where $\text{Proj}_{\mathbf{s}}$ is the same orthogonal projector as in (\ref{equ25}) but at point $\mathbf{s}$. Taking the Euclidean gradient of $f_{\mathbf{s}}$, one can derive from (\ref{equ34}) that
\begin{equation}
\label{equ36}
\begin{split}
\mathbf{Grad}f_{\mathbf{s}}&= \frac{\sum_{i=0}^{N_{t}-1}2(\mathbf{s}_{k}^{H}\mathbf{\Psi}_{i}^{H}\mathbf{s}_{k}\mathbf{\Psi}_{i}\mathbf{s}_{k}+\mathbf{s}^{H}\mathbf{\Psi}_{i}\mathbf{s}\mathbf{\Psi}_{i}^{H}\mathbf{s})\lvert\tilde{\mathbf{s}}^{H}\mathbf{s}_{k}\rvert^{2}}{\lvert\mathbf{s}_{k}^{H}\tilde{\mathbf{s}}\rvert^{4}}\\
&-\frac{\sum_{i=0}^{N_{t}-1}2(\mathbf{s}_{k}^{H}\mathbf{\Psi}_{i}\mathbf{s}_{k}\mathbf{s}_{k}^{H}\mathbf{\Psi}_{i}^{H}\mathbf{s}_{k})(\tilde{\mathbf{s}}\tilde{\mathbf{s}}^{H}\mathbf{s})}{\lvert\mathbf{s}_{k}^{H}\tilde{\mathbf{s}}\rvert^{4}}
\end{split}.
\end{equation}

The Riemannian Hessian map of $f_{\mathbf{s}}$ can be calculated in the same fashion as (\ref{equ27}). The directional derivative in the calculation is
\begin{equation}
\label{equ37}
\begin{split}
D{\mathbf{Grad}}f_{{\mathbf{s}}}(\mathbf{\xi}_{\mathbf{x}})&=\frac{\sum_{i=0}^{N_{t}-1}2(\mathbf{\zeta}_{1}(i)+\mathbf{\zeta}_{2}(i)-\mathbf{\zeta}_{3}(i))}{\gamma^{2}}\\
&\quad+\frac{\sum_{i=0}^{N_{t}-1}2(\mathbf{\zeta}_{4}(i)-\mathbf{\zeta}_{5}(i))}{\gamma^{4}}, \end{split}
\end{equation}
where we have
\begin{equation}
\label{equ38}
\begin{split}
\gamma &= \lvert \mathbf{s}_{k}^{H}\tilde{\mathbf{s}}\rvert^{2}\\
\mathbf{\zeta}_{1}(i) &= \mathbf{\xi}_{\mathbf{x}}^{H}\mathbf{\Psi}_{i}^{H}\mathbf{s}_{k}\mathbf{\Psi}_{i}\mathbf{s}_{k}\gamma+\mathbf{s}_{k}^{H}\mathbf{\Psi}_{i}^{H}\mathbf{\xi}_{\mathbf{x}}\mathbf{\Psi}_{i}\mathbf{s}_{k}\gamma+\mathbf{s}_{k}^{H}\mathbf{\Psi}_{i}^{H}\mathbf{s}_{k}\mathbf{\Psi}_{i}\mathbf{\xi}_{\mathbf{x}}\gamma\\
& \quad+\mathbf{s}_{k}^{H}\mathbf{\Psi}_{i}^{H}\mathbf{s}_{k}\mathbf{\Psi}_{i}\mathbf{s}_{k}\tilde{\mathbf{s}}^{H}\mathbf{\xi}_{\mathbf{x}}\mathbf{s}_{k}^{H}\tilde{\mathbf{s}}+\mathbf{s}_{k}^{H}\mathbf{\Psi}_{i}^{H}\mathbf{s}_{k}\mathbf{\Psi}_{i}\mathbf{s}_{k}\tilde{\mathbf{s}}^{H}\mathbf{s}_{k}\mathbf{\xi}_{\mathbf{x}}^{H}\tilde{\mathbf{s}}\\
\mathbf{\zeta}_{2}(i) & = \mathbf{\xi}_{\mathbf{x}}^{H}\mathbf{\Psi}_{i}\mathbf{s}_{k}\mathbf{\Psi}_{i}^{H}\mathbf{s}_{k}\gamma+\mathbf{s}_{k}^{H}\mathbf{\Psi}_{i}\mathbf{\xi}_{\mathbf{x}}\mathbf{\Psi}_{i}^{H}\mathbf{s}_{k}\gamma+\mathbf{s}_{k}^{H}\mathbf{\Psi}_{i}\mathbf{s}_{k}\mathbf{\Psi}_{i}^{H}\mathbf{\xi}_{\mathbf{x}}\gamma\\
&\quad +\mathbf{s}_{k}^{H}\mathbf{\Psi}_{i}\mathbf{s}_{k}\mathbf{\Psi}_{i}^{H}\mathbf{s}_{k}\tilde{\mathbf{s}}^{H}\mathbf{\xi}_{\mathbf{x}}\mathbf{s}_{k}^{H}\tilde{\mathbf{s}}+\mathbf{s}_{k}^{H}\mathbf{\Psi}_{i}\mathbf{s}_{k}\mathbf{\Psi}_{i}^{H}\mathbf{s}_{k}\tilde{\mathbf{s}}^{H}\mathbf{s}_{k}\mathbf{\xi}_{\mathbf{x}}^{H}\tilde{\mathbf{s}}\\
\mathbf{\zeta}_{3}(i) & =\mathbf{\xi}_{\mathbf{x}}^{H}\mathbf{\Psi}_{i}\mathbf{s}_{k}\mathbf{s}_{k}^{H}\mathbf{\Psi}_{i}^{H}\mathbf{s}_{k}\tilde{\mathbf{s}}\tilde{\mathbf{s}}^{H}\mathbf{s}_{k}+\mathbf{s}_{k}^{H}\mathbf{\Psi}_{i}\mathbf{\xi}_{\mathbf{x}}\mathbf{s}_{k}^{H}\mathbf{\Psi}_{i}^{H}\mathbf{s}_{k}\tilde{\mathbf{s}}\tilde{\mathbf{s}}^{H}\mathbf{s}_{k}\\
& \quad+\mathbf{s}_{k}^{H}\mathbf{\Psi}_{i}\mathbf{s}_{k}\mathbf{\xi}_{\mathbf{x}}^{H}\mathbf{\Psi}_{i}^{H}\mathbf{s}_{k}\tilde{\mathbf{s}}\tilde{\mathbf{s}}^{H}\mathbf{s}_{k}+\mathbf{s}_{k}^{H}\mathbf{\Psi}_{i}\mathbf{s}_{k}\mathbf{s}_{k}^{H}\mathbf{\Psi}_{i}^{H}\mathbf{\xi}_{\mathbf{x}}\tilde{\mathbf{s}}\tilde{\mathbf{s}}^{H}\mathbf{s}_{k}\\
&\quad+\mathbf{s}_{k}^{H}\mathbf{\Psi}_{i}\mathbf{s}_{k}\mathbf{s}_{k}^{H}\mathbf{\Psi}_{i}^{H}\mathbf{s}_{k}\tilde{\mathbf{s}}\tilde{\mathbf{s}}^{H}\mathbf{\xi}_{\mathbf{x}} \\
\mathbf{\zeta}_{4}(i) & = (\mathbf{s}_{k}^{H}\mathbf{\Psi}_{i}\mathbf{s}_{k}\mathbf{s}_{k}^{H}\mathbf{\Psi}_{i}^{H}\mathbf{s}_{k})(\tilde{\mathbf{s}}\tilde{\mathbf{s}}^{H}\mathbf{s}_{k})(\mathbf{\xi}_{\mathbf{x}}^{H}\tilde{\mathbf{s}}\tilde{\mathbf{s}}^{H}\mathbf{s}_{k}\mathbf{s}_{k}^{H}\tilde{\mathbf{s}}\tilde{\mathbf{s}}^{H}\mathbf{s}_{k}\\
&\quad +\mathbf{s}_{k}^{H}\tilde{\mathbf{s}}\tilde{\mathbf{s}}^{H}\mathbf{\xi}_{\mathbf{x}}\mathbf{s}_{k}^{H}\tilde{\mathbf{s}}\tilde{\mathbf{s}}^{H}\mathbf{s}_{k}+\mathbf{s}_{k}^{H}\tilde{\mathbf{s}}\tilde{\mathbf{s}}^{H}\mathbf{s}_{k}\mathbf{\xi}_{\mathbf{x}}^{H}\tilde{\mathbf{s}}\tilde{\mathbf{s}}^{H}\mathbf{s}_{k}\\
&\quad+\mathbf{s}_{k}^{H}\tilde{\mathbf{s}}\tilde{\mathbf{s}}^{H}\mathbf{s}_{k}\mathbf{s}_{k}^{H}\tilde{\mathbf{s}}\tilde{\mathbf{s}}^{H}\mathbf{\xi}_{\mathbf{x}})\\
\mathbf{\zeta}_{5}(i) & =
\gamma(\mathbf{s}_{k}^{H}\mathbf{\Psi}_{i}^{H}\mathbf{s}_{k}\mathbf{\Psi}_{i}\mathbf{s}_{k}+\mathbf{s}_{k}^{H}\mathbf{\Psi}_{i}\mathbf{s}_{k}\mathbf{\Psi}_{i}^{H}\mathbf{s}_{k})\\
&\quad(\mathbf{\xi}_{\mathbf{x}}^{H}\tilde{\mathbf{s}}\tilde{\mathbf{s}}^{H}\mathbf{s}_{k}\mathbf{s}_{k}^{H}\tilde{\mathbf{s}}\tilde{\mathbf{s}}^{H}\mathbf{s}_{k}+\mathbf{s}_{k}^{H}\tilde{\mathbf{s}}\tilde{\mathbf{s}}^{H}\mathbf{\xi}_{\mathbf{x}}\mathbf{s}_{k}^{H}\tilde{\mathbf{s}}\tilde{\mathbf{s}}^{H}\mathbf{s}_{k}\\
&\quad+\mathbf{s}_{k}^{H}\tilde{\mathbf{s}}\tilde{\mathbf{s}}^{H}\mathbf{s}_{k}\mathbf{\xi}_{\mathbf{x}}^{H}\tilde{\mathbf{s}}\tilde{\mathbf{s}}^{H}\mathbf{s}_{k}+\mathbf{s}_{k}^{H}\tilde{\mathbf{s}}\tilde{\mathbf{s}}^{H}\mathbf{s}_{k}\mathbf{s}_{k}^{H}\tilde{\mathbf{s}}\tilde{\mathbf{s}}^{H}\mathbf{\xi}_{\mathbf{x}}).
\end{split}
\end{equation}
Then the Riemannian Hessian can be calculated by
\begin{equation}
\label{equ39}
\begin{aligned}
    \mathbf{Hess}f_{\mathbf{s}}(\mathbf{\xi}_{\mathbf{x}}) 
    & = \text{Proj}_{\mathbf{s}_{k}}(D{\mathbf{Grad}}f_{\mathbf{s}}(\mathbf{\xi}_{\mathbf{x}})) \\
    & - \mathop{Re}\left\{\mathbf{Grad}f_{\mathbf{s}}\odot\mathbf{s}_{k}\right\}\odot \mathbf{\xi}_{\mathbf{x}}.
\end{aligned}
\end{equation}

Again, we have derived all the ingredients of $h_{k}(\mathbf{\xi}_{\mathbf{x}})$ and can obtain the best step $\mathbf{\xi}_{\mathbf{x}}$ by solving the sub-problem (\ref{equ33}) with TCG. Then the approximation improvement should be evaluated at each tentative updated point $\textbf{R}_{\mathbf{s}_{k}}(\mathbf{\xi}_{\mathbf{x}}^{\star})$ with the ratio
\begin{equation}
\label{equ40}
\rho_{k} = \frac{f_{\mathbf{s}}(\mathbf{s}_{k})-f_{\mathbf{s}}(\mathbf{s}_{k+1})}{h_{k}(\mathbf{0})-h_{k}(\mathbf{\xi}_{\mathbf{x}}^{\star})}.
\end{equation}

Next, we formulate the RTR algorithm to solve (\ref{equ10.2}) as follows:
\begin{figure}[H]
\begin{tabular}{p{8.5cm}}
\hline
\textbf{Algorithm 2: RTR method to solve (\ref{equ10.2})} \\
\hline
\begin{enumerate}
\item \textbf{Parameters}: maximum trust region radius $\bar{\Delta}$, threshold $\bar{\rho}\in(0,\frac{1}{4})$, user-determined threshold $\epsilon$.
  \item \textbf{Input}: $\mathbf{s}_{0}\in \mathcal{M}, \Delta_{0}\in (0,\bar{\Delta}]$
  \item \textbf{For} $k=0,1,2,\cdots$
  \begin{enumerate}
  \item Calculate the Riemannian gradient at $\mathbf{s}_{k}$ using (\ref{equ35}) and (\ref{equ36}), formalize the Hessian map by (\ref{equ37}), (\ref{equ38}) and (\ref{equ39}).
  \item Use TCG method to solve the sub-problem (\ref{equ33}), yielding $\mathbf{\xi}_{\mathbf{x}}^{\star}$.
  \item Compute $\rho_{k}$ by (\ref{equ40}).
  \item Accept or reject the tentative candidate:
  \begin{equation}
    \label{equ41}
    \mathbf{s}_{k+1}=\left\{\begin{matrix} \textbf{R}_{\mathbf{s}_{k}}(\mathbf{\xi}_{\mathbf{x}}^{\star}) & \text{if}\, \rho_{k}>\bar{\rho}\, \text{(accept)} \\ \mathbf{s}_{k} & \text{otherwise (reject)} \end{matrix}\right.
  \end{equation}
  \item Update the trust-region radius:
   \begin{equation}
    \label{equ42}
    \Delta_{k+1}=\left\{\begin{matrix} \frac{1}{4}\Delta_{k} & \text{if}\, \rho_{k}<\frac{1}{4} \\ \mathop{min}(2\Delta_{k},\bar{\Delta}) & \begin{split} & \text{if}\, \rho_{k}>\frac{3}{4}\, \text{and}\\ &\|\mathbf{\xi}_{\mathbf{x}}^{\star}\|_{\mathbf{s}_{k}}=\Delta_{k} \end{split}  \\ \Delta_{k} & \text{otherwise} \end{matrix}\right.
  \end{equation}
  \item Until $\|\mathbf{grad}f_{\mathbf{s}}\|_{\mathbf{s}_{k+1}} \le \epsilon$.
  \end{enumerate}
  \item \textbf{Output}: output $\mathbf{s}_{k+1}$.
\end{enumerate}\\
\hline
\end{tabular}
\end{figure}

\subsection{Algorithm summary and convergence analysis}
As stated in section \ref{section: III}, our problem is a two-step optimization. Thus to obtain a final optimized transmit sequence, our algorithm alternatively solves sub-problems (\ref{equ10.1}) and (\ref{equ10.2}). When optimizing the target Doppler steering vector $\tilde{\mathbf{s}}$, the transmit sequence $\mathbf{s}$ is considered to be a priori parameter obtained from the last optimization of (\ref{equ10.2}). Likewise, when optimizing the transmit sequence $\mathbf{s}$, the target Doppler steering vector is assumed to be obtained.

Since our algorithm optimizes the transmit sequence for the worst case with the RTR method, we name the algorithm Worst-case Riemannian Trust Region (WRTR) method.

For the convergence of the proposed algorithm, we give the following theorems:

$\textit{Theorem}\, \textit{2}$ :
\begin{theorem}
The global convergence of Algorithm 1 is guaranteed, i.e. 
\begin{equation}
\label{equ43}
\lim_{k\to\infty}\mathbf{grad}f_{{\tilde{\mathbf{s}}}}({\tilde{\mathbf{s}}}_{k})=0.
\end{equation}
The proof of the theorem can be found in Appendix \ref{app3}.  
\end{theorem}

$\textit{Theorem}\, \textit{3}$ :
\begin{theorem}
Let $\mathbf{v}$ be a non-degenerate local minimizer of $f_{\mathbf{s}}$, i.e., $\mathbf{grad}f_{\mathbf{s}}(\mathbf{v})=0$ and $\mathbf{Hess}f(\mathbf{v})$ is positive definite, then there exists a neighborhood $V$ of $\mathbf{v}$ such that, for all $\mathbf{s}_{0}\in V$, the sequence $\{\mathbf{s}_{k}\}$ generated by Algorithm 2 converges to $\mathbf{v}$.
\end{theorem}

The proof of the theorem can be found in Appendix \ref{app4}.  

Although algorithm 1 and algorithm 2  are guaranteed to converge with an infinite number of iterations, they usually reach the following termination conditions in a bounded number of iterations \cite{boumal2019global}
\begin{equation}
\label{equ44}
\|\mathbf{grad}f(\mathbf{s}_{k})\|\le \epsilon_{g} \,\text{and}\, \lambda_{min}(\mathbf{Hess}f_{{\mathbf{s}}})\ge \epsilon_{h}.
\end{equation}
where $\epsilon_{g}$ and $\epsilon_{h}$ are user-determined thresholds, $f$ represents $f_{\mathbf{s}}$ or $f_{\tilde{\mathbf{s}}}$. 
We should point out that the theoretical convergence of the proposed WRTR remains open for debate. However, we analyzed it in the following simulation section and found that WRTR converged only after a few iterations.

\section{Simulation results and discussions}
In this section, we perform several numerical experiments to demonstrate the effectiveness and superiority of our proposed algorithms. Firstly, we evaluate the convergence performance of our method by varying different parameters. Next, we compare the achieved SCNR of our developed WRTR method with that of other widely used algorithms. Furthermore, we present and analyze the corresponding STAFs of different algorithms to illustrate their interference suppression capabilities.

\subsection{Convergence performance}
In this subsection, we focus on studying the convergence performance of our proposed algorithm. Specifically, we set the code length $N$ and normalized Doppler bins to 64. The range bins occupied by the clutter scatters are defined as $r_k \in \left\{ 11,12,....,30 \right\}$. The corresponding normalized Doppler of these scatters is assumed to vary from 25 to 26. Additionally, the complex amplitude $\varrho_k$ of the clutter is set to 10 dB, and the Lagrange multiplier $\lambda$ is set to 100. Furthermore, we primarily consider the phase error resulting from the Doppler frequency estimation. Consequently, the threshold $\epsilon$ can be obtained as
\begin{equation}
    \epsilon = \underset{v \in v_{\mathcal{I}}} {\operatorname{max}} \left\| {\mathbf{p}}(v) - {\mathbf{p}}(v_{t}) \right\|^2,
\end{equation}
where $v_I$ denotes the estimated Doppler frequency set. Here we set $v_{t} \in \left[ -0.1, 0.1 \right]$.
For the convergence condition, we stop the WRTR algorithm once the maximum iterations or the norm of the gradient (NoG) reaches the predefined parameters. In this experiment, the maximum number of iterations is set to $100$, and the tolerance of the NoGs is $1 \times 10^{-9}$.

\begin{figure}
    \centering
    \includegraphics[width=5cm]{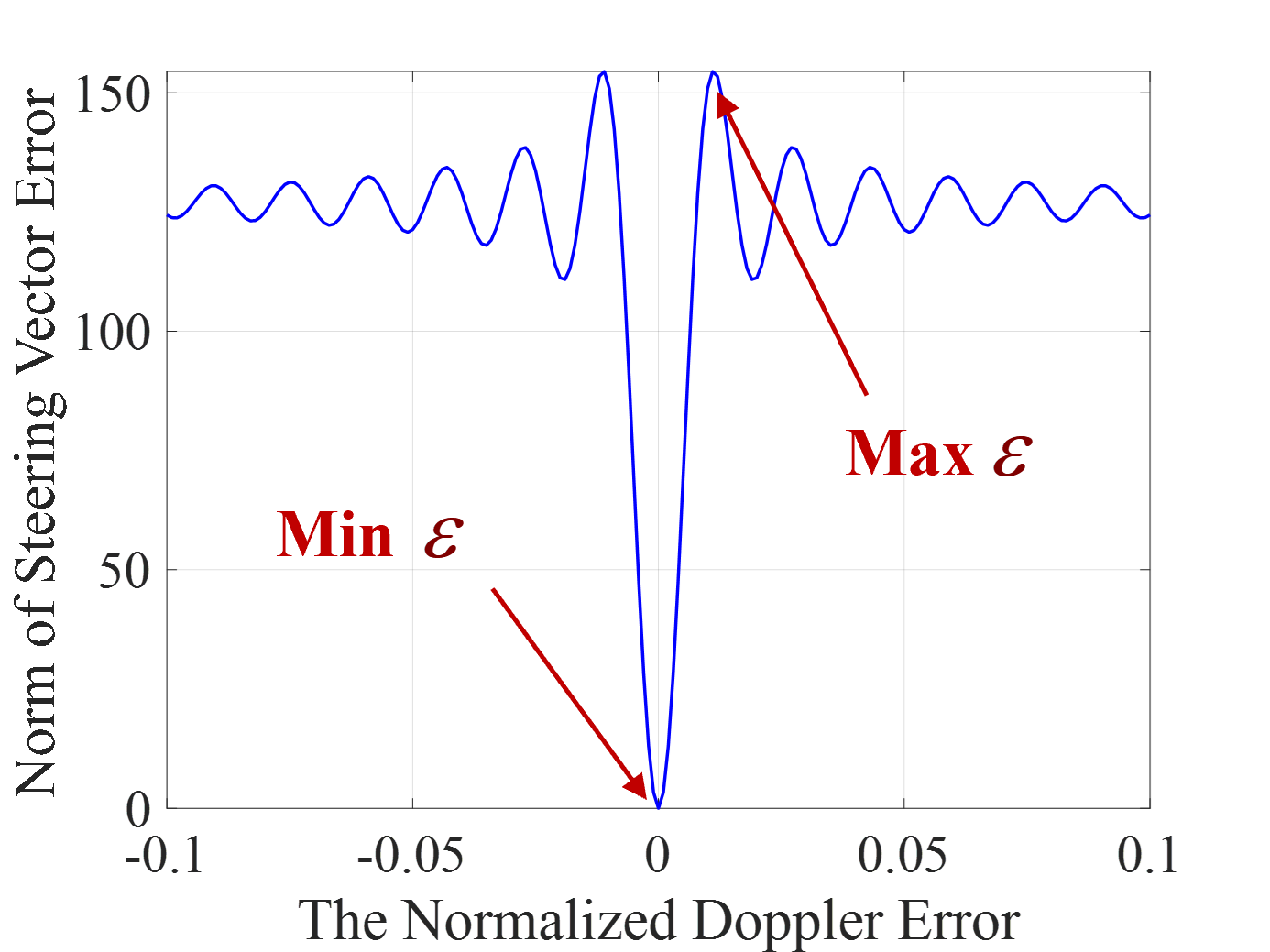}
    \caption{The Norm of Steering Vector Error}
    \label{epsilon}
\end{figure}

\begin{figure}
	\centering
	\subfigure[cost values with the iterations]{
		\begin{minipage}[t]{0.5\linewidth}
			\centering
			\includegraphics[width=4.5cm]{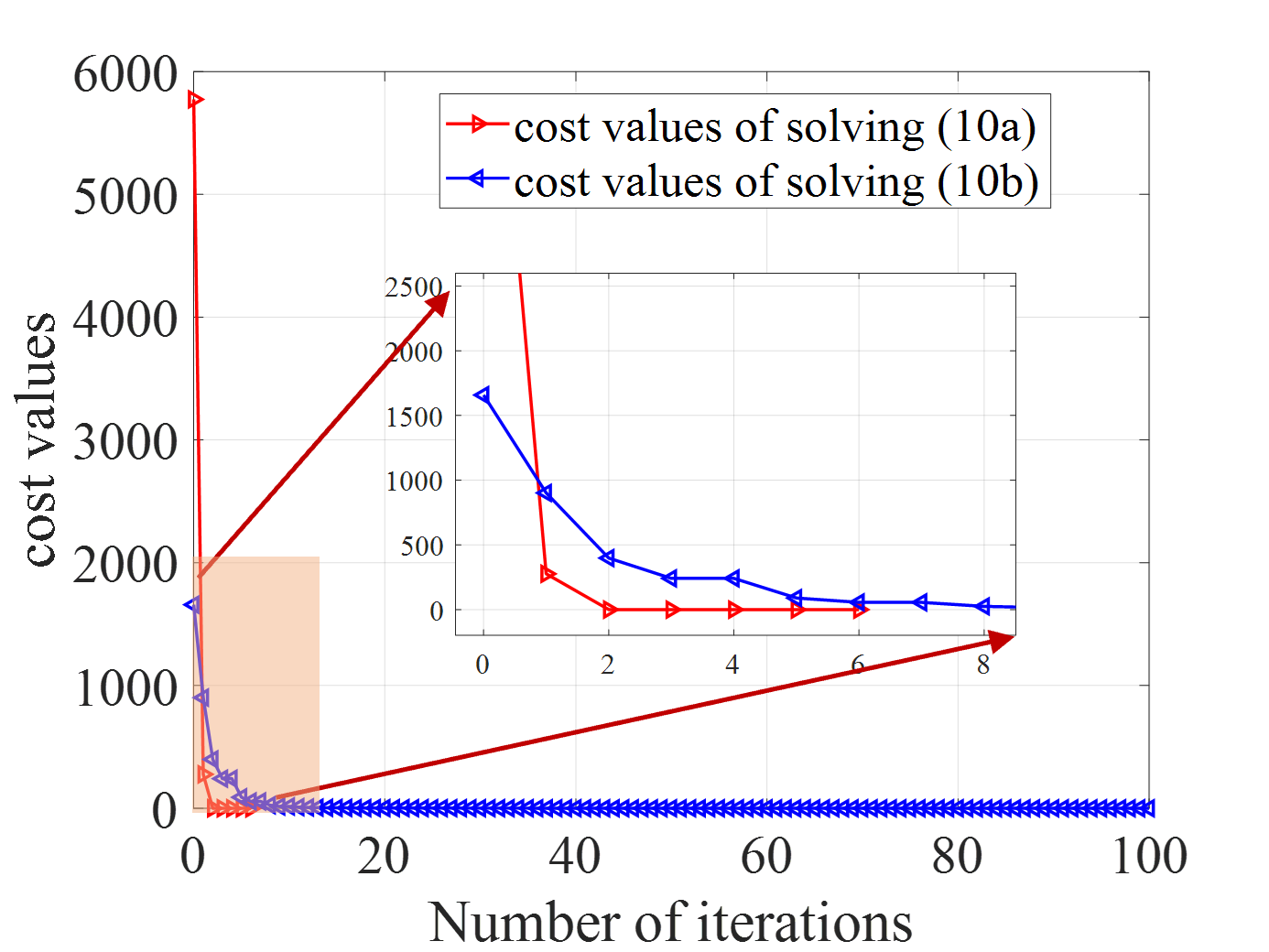}
		\end{minipage}
	}%
	\subfigure[NoGs with the iterations]{
		\begin{minipage}[t]{0.5\linewidth}
			\centering
			\includegraphics[width=4.5cm]{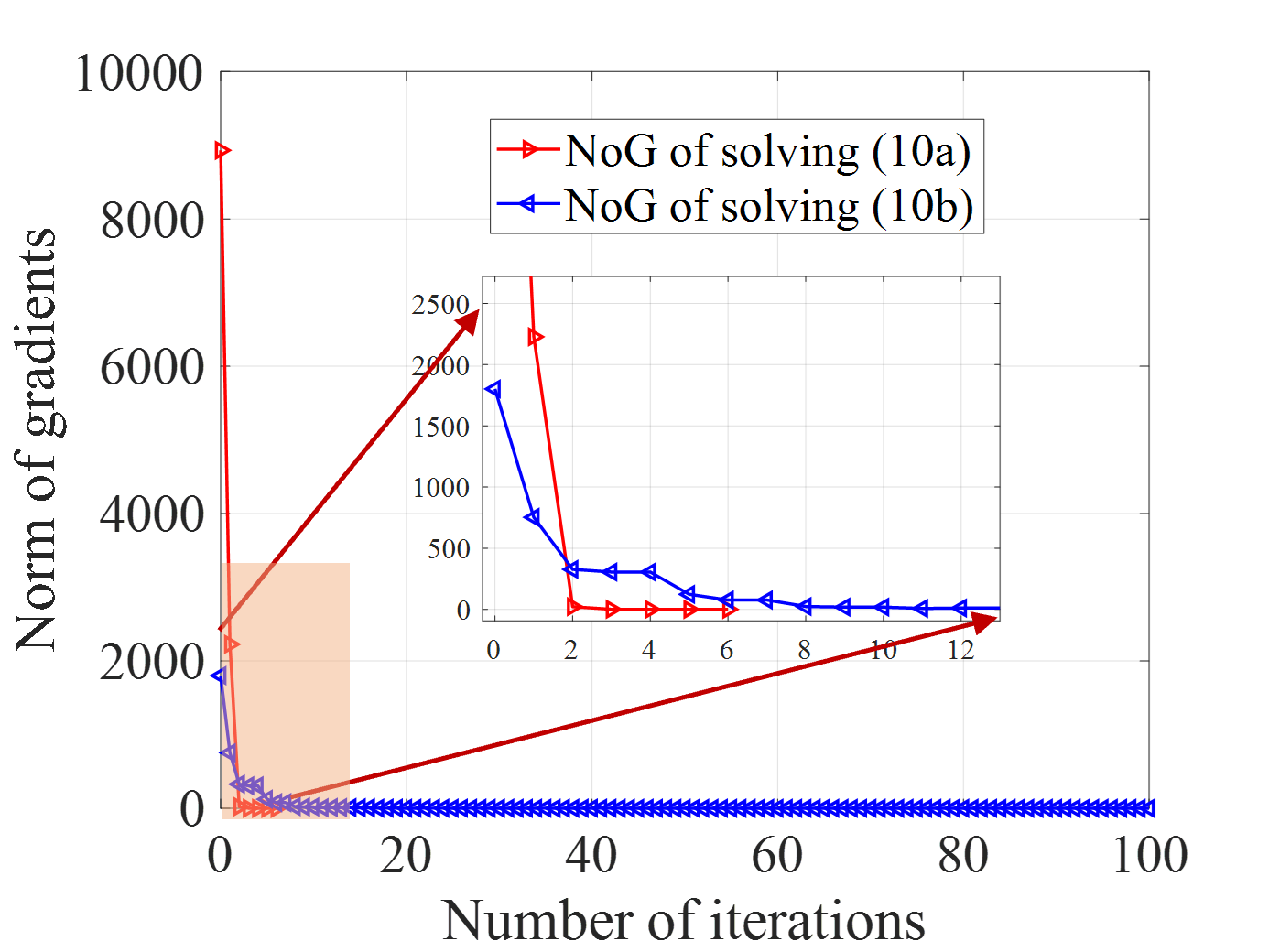}
		\end{minipage}%
	}%
	\caption{the cost values and norms of gradients versus the number of iterations.}
	\label{cost_grad_update}
\end{figure}

We begin by exploring the impact of Doppler error on the threshold value $\epsilon$. As depicted in Figure \ref{epsilon}, when the Doppler error is small, meaning it is close to the actual target Doppler frequency, the threshold $\epsilon$ assumes a larger value. As the error gradually increases, the threshold value stabilizes and converges in an oscillatory manner. Notably, when the Doppler error is zero, the threshold value should also be zero.
In Fig. \ref{cost_grad_update}, we present the cost values and NoGs obtained while solving the two sub-problems \eqref{equ10.1} and \eqref{equ10.2} during the initial iteration. As shown in Fig \ref{cost_grad_update}(a), the cost function values for both sub-problems steadily decrease. Specifically, the first sub-problem is solved within 10 iterations, while the second sub-problem reaches convergence after the maximum iterations are reached.
In Fig. \ref{cost_grad_update}(b), the NoGs do not exhibit a monotonic decrease as the algorithm progresses. This behavior can be attributed to the non-convex nature of the sub-problems. These results collectively demonstrate the efficacy of our proposed algorithm in addressing problem \eqref{equ10}.

\begin{figure}
	\centering
	\subfigure[Spectrum of solving \eqref{equ10.1}]{
		\begin{minipage}[t]{0.5\linewidth}
			\centering
			\includegraphics[width=4.5cm]{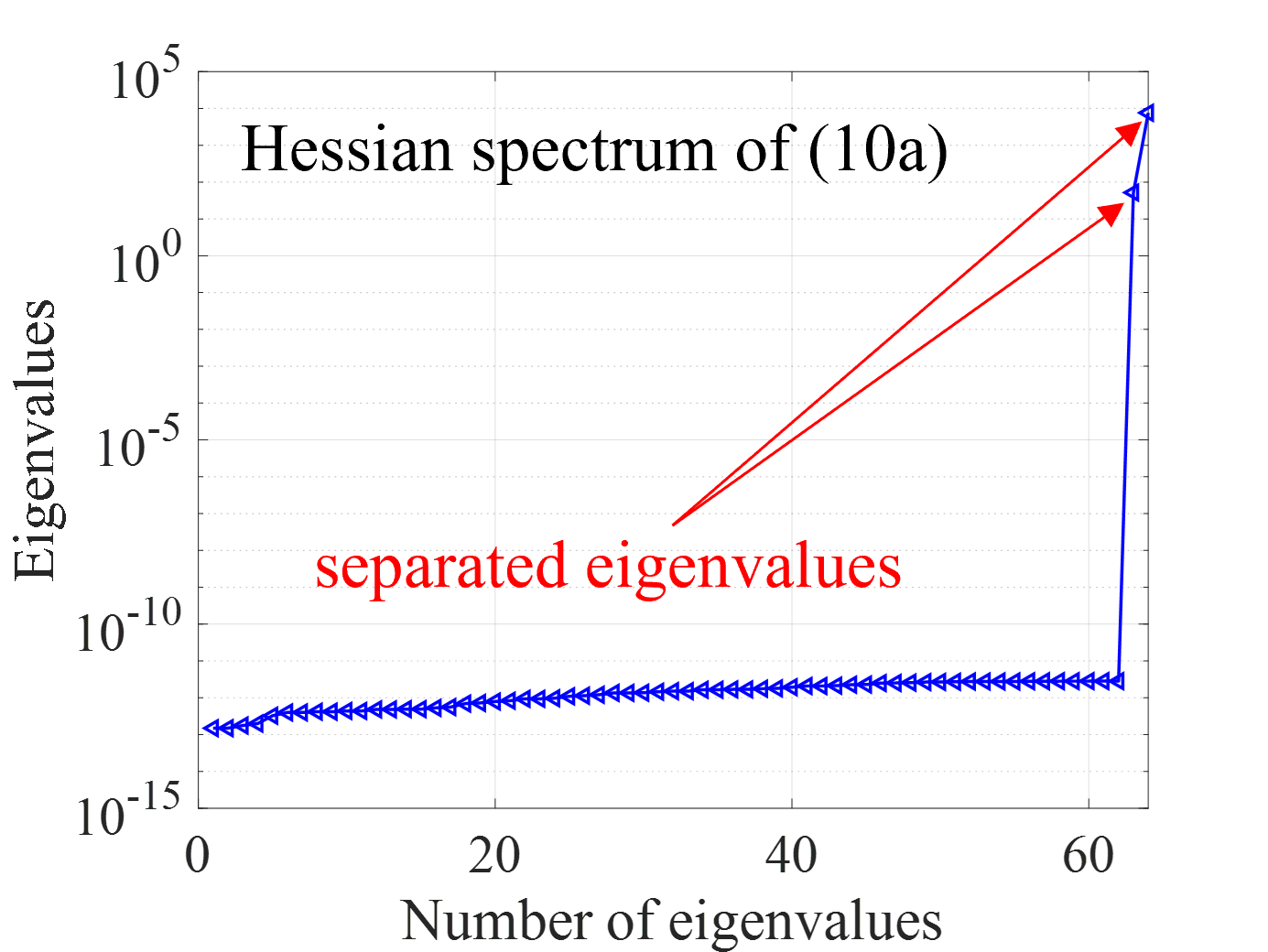}
		\end{minipage}
	}%
	\subfigure[Spectrum of solving \eqref{equ10.1}]{
		\begin{minipage}[t]{0.5\linewidth}
			\centering
			\includegraphics[width=4.5cm]{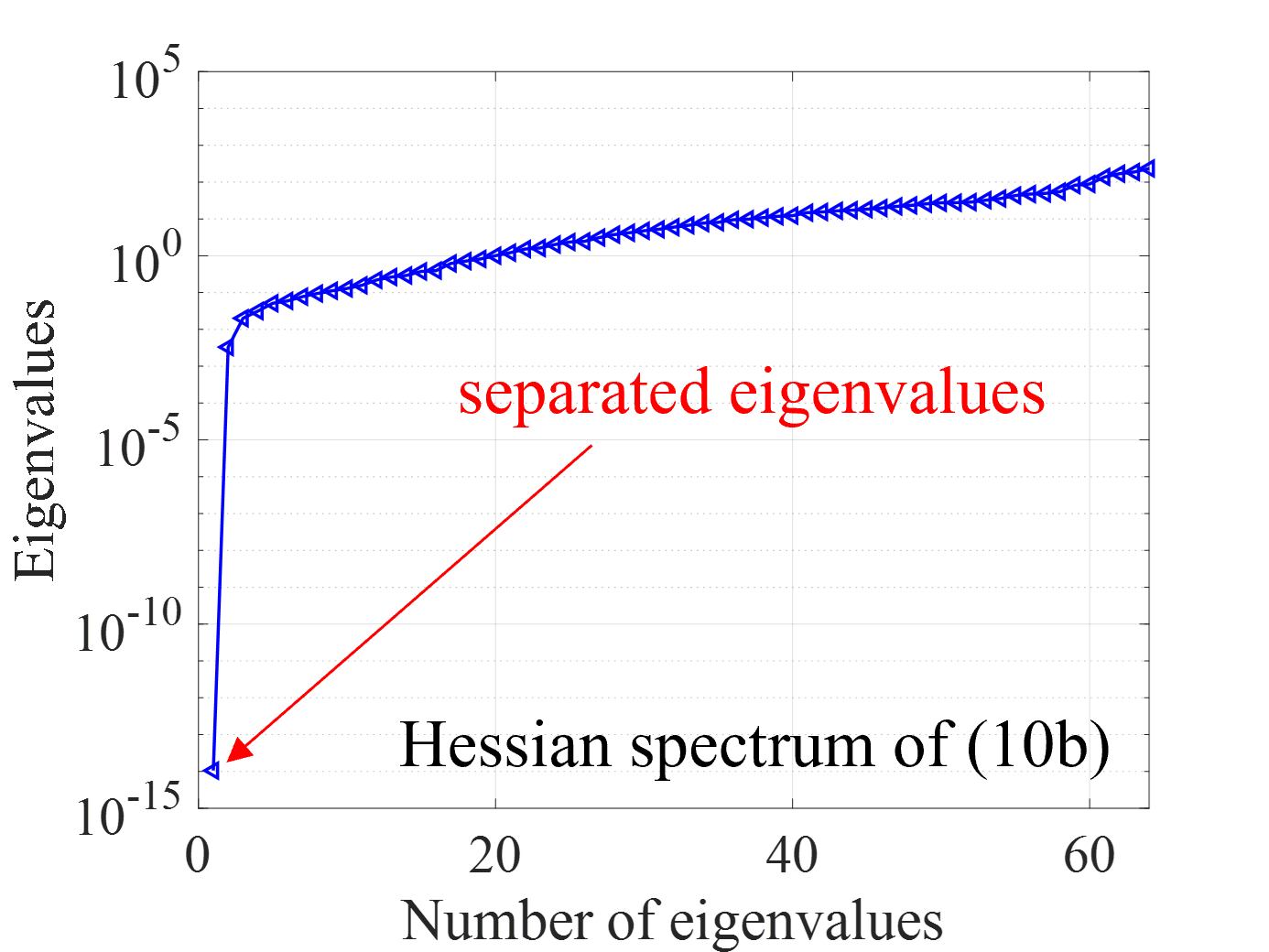}
		\end{minipage}%
	}%
	\caption{the Hessian spectrums of solving two sub-problems \eqref{equ10.1} and \eqref{equ10.2}.}
	\label{HessianSpectrum}
\end{figure}
To further validate the local minima nature of the iterative points obtained by the proposed algorithms, we calculate the Riemannian Hessian spectrum of these points and present the results in Fig. \ref{HessianSpectrum}. 
It is worth noting that each eigenvalue of the Riemannian Hessian is non-negative, indicating that the corresponding Riemannian Hessian matrices of critical points are semi-definite positive and effectively avoid saddle points. 
Fig. \ref{HessianSpectrum} illustrates that the spectrum of Riemannian Hessian matrices exhibits a linear distribution, making it easy to identify the largest eigenvalue of the Riemannian Hessian. 
The aforementioned discussions provide confirmation that critical points with Riemannian gradients converging to zero are indeed local minima of the sub-problems.

\begin{figure}
	\centering
	\subfigure[the influence of Doppler error]{
		\begin{minipage}[t]{0.5\linewidth}
			\centering
			\includegraphics[width=4.5cm]{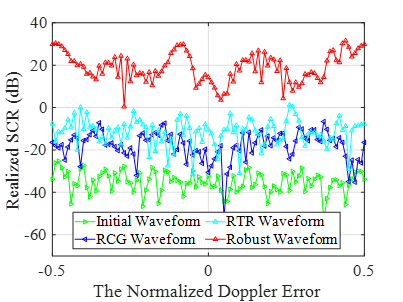}
		\end{minipage}
	}%
	\subfigure[the influence of Doppler error]{
		\begin{minipage}[t]{0.5\linewidth}
			\centering
			\includegraphics[width=4.5cm]{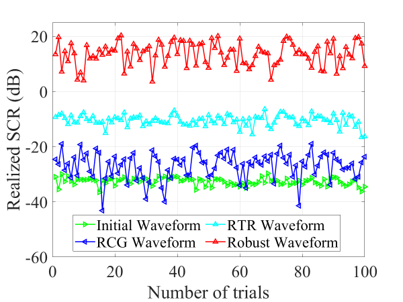}
		\end{minipage}%
	}%
	\caption{the realized SCRs value versus different errors.}
	\label{realized SCR}
\end{figure}

We further investigate the performance of our proposed algorithm in terms of Signal-to-Clutter Ratio (SCR). In Fig.\ref{realized SCR}(a), we compare the realized SCR values obtained by different algorithms. To simplify the following discussion, we represent the waveforms designed using the Riemannian conjugated gradient (RCG) algorithm and Riemannian trust region (RTR) algorithm as RCG waveform and RTR waveform, respectively.
As shown in the figure, both the RCG and RTR methods achieve a significant improvement of 20dB in SCR compared to the initial waveform. In contrast, traditional waveform design methods that do not consider Doppler errors result in much smaller realized SCRs.
In Fig.\ref{realized SCR}(b), we further investigate the impact of random steering vector $\boldsymbol{p}\left( v_t \right)$ on the realized SCR values. In this experiment, the phase value of each element in the steering vector is generated from a standard uniform distribution.
Based on the results of 100 Monte Carlo experiments, it is evident that the robust waveform designed by our proposed algorithm achieves the highest SCR. This demonstrates its robustness to phase errors in the target steering vector. Furthermore, comparing the experimental results, we also observe that the realized SCR in Fig.\ref{realized SCR}(b) is significantly lower than that in Fig.\ref{realized SCR}(a). This indicates that the influence of a random steering vector on SCR is much larger than that of random phase errors.

\subsection{Interference suppression performance}
\begin{figure}
	\centering
	\subfigure[STAF of the initial waveform]{
		\begin{minipage}[t]{0.5\linewidth}
			\centering
			\includegraphics[width=4.5cm]{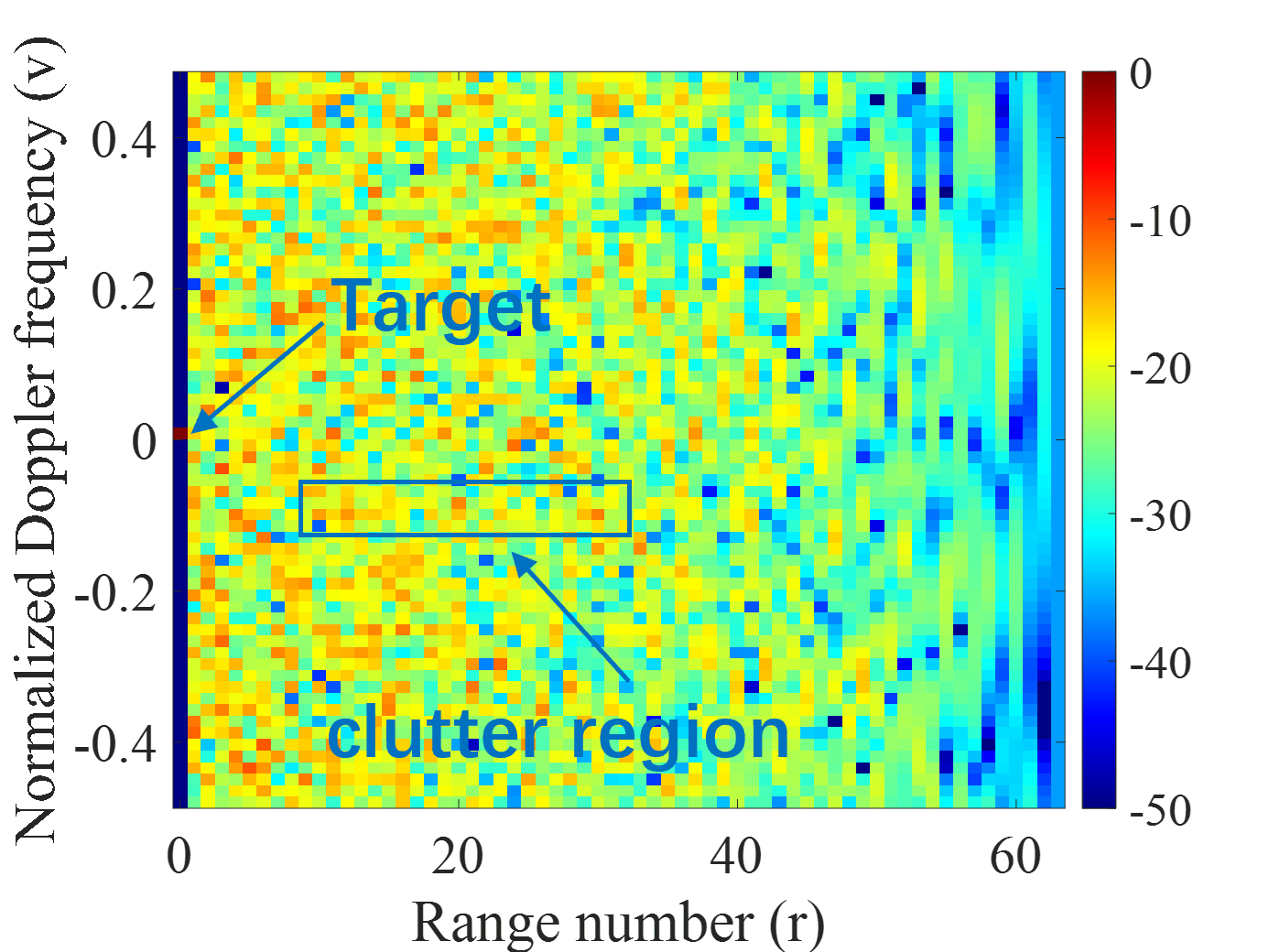}
		\end{minipage}
	}%
	\subfigure[STAF of the RCG waveform]{
		\begin{minipage}[t]{0.5\linewidth}
			\centering
			\includegraphics[width=4.5cm]{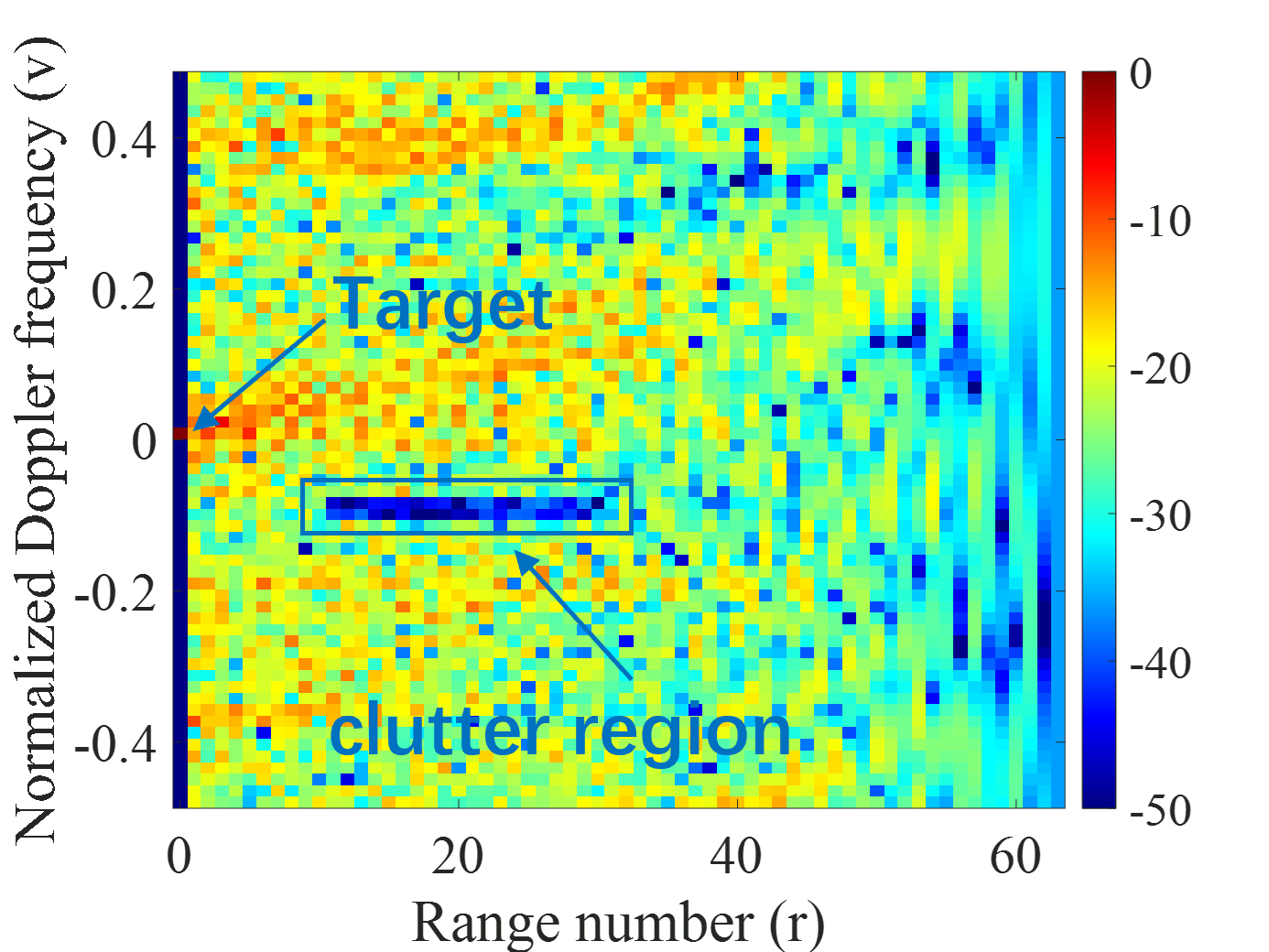}
		\end{minipage}%
	}%

 	\subfigure[STAF of the RTR waveform]{
		\begin{minipage}[t]{0.5\linewidth}
			\centering
			\includegraphics[width=4.5cm]{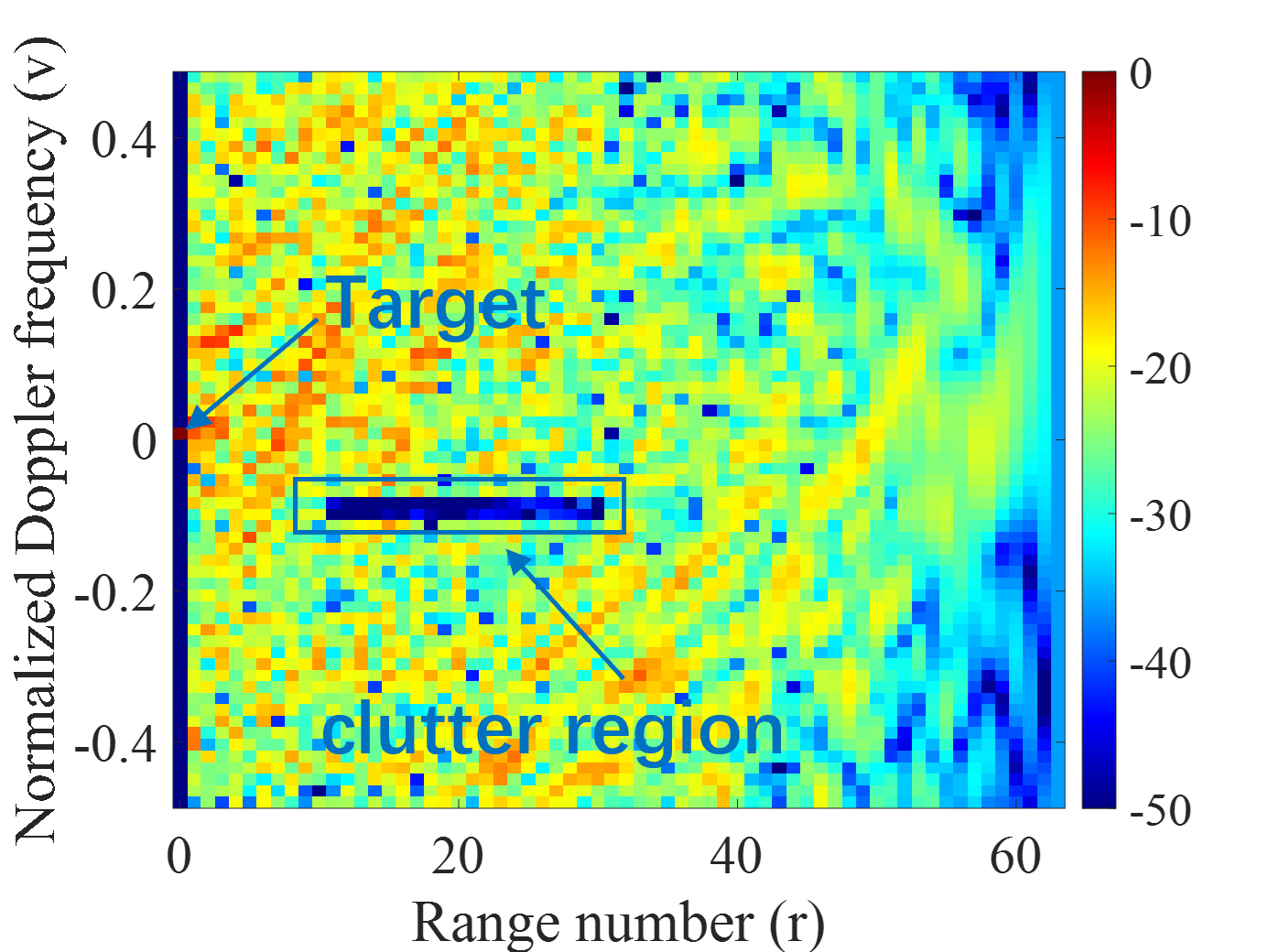}
		\end{minipage}
	}%
	\subfigure[STAF of the robust waveform]{
		\begin{minipage}[t]{0.5\linewidth}
			\centering
			\includegraphics[width=4.5cm]{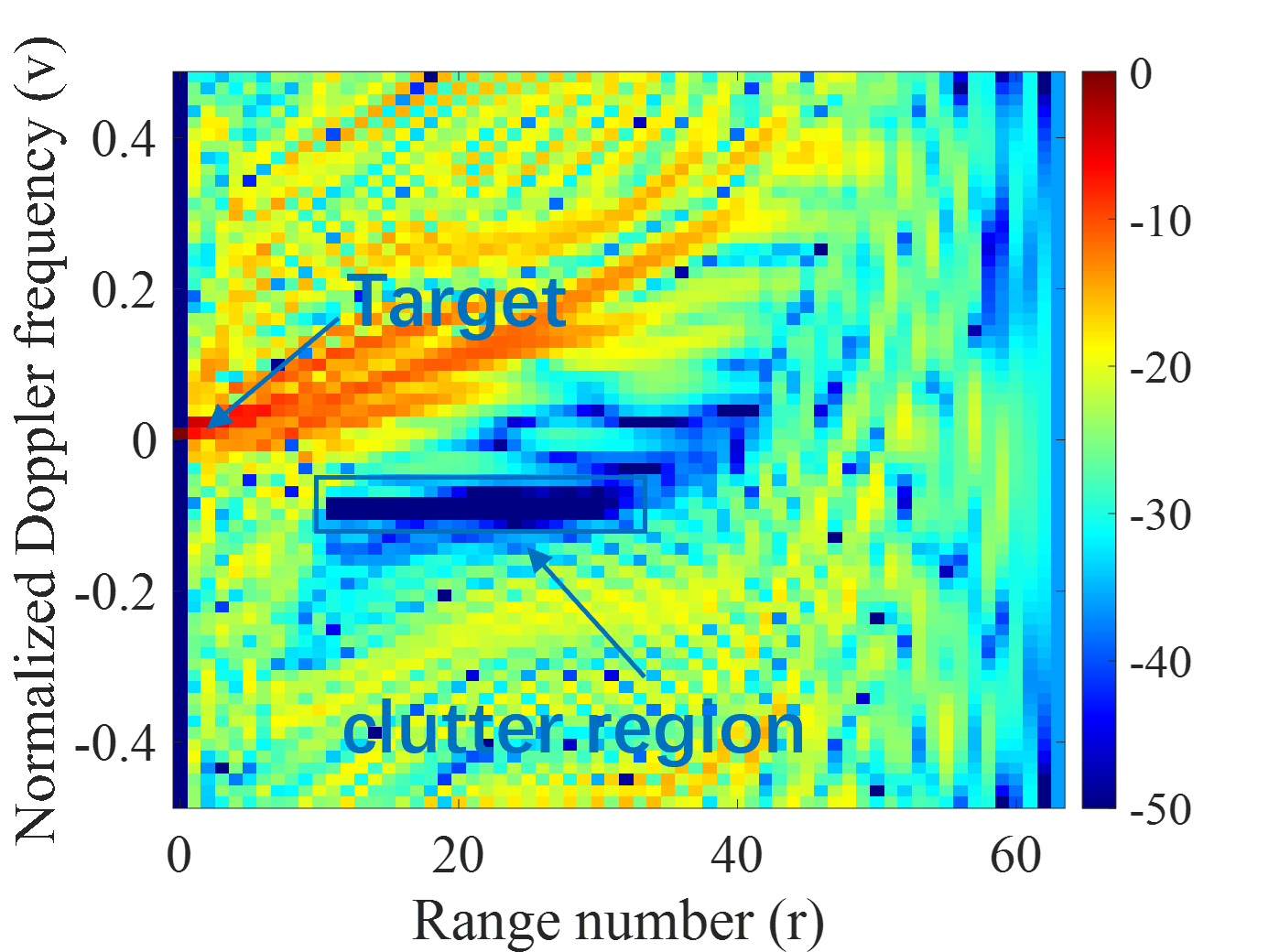}
		\end{minipage}%
	}%

  	\subfigure[Doppler cut of $l=25$]{
		\begin{minipage}[t]{0.5\linewidth}
			\centering
			\includegraphics[width=4.5cm]{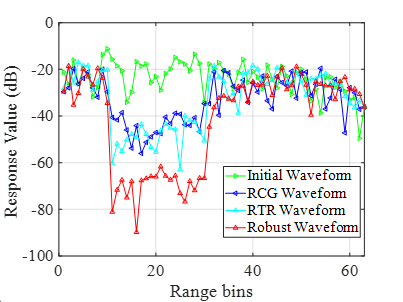}
		\end{minipage}
	}%
	\subfigure[Doppler cut of $l=26$]{
		\begin{minipage}[t]{0.5\linewidth}
			\centering
			\includegraphics[width=4.5cm]{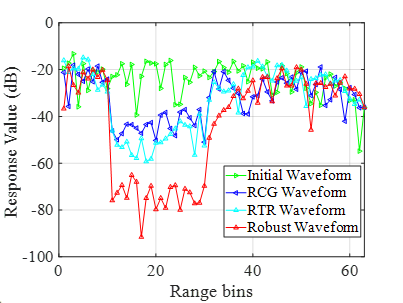}
		\end{minipage}%
	}%
	\caption{the realized STAFs and Doppler cuts corresponding to different waveforms.}
	\label{shaping result}
\end{figure}

In this subsection, we analyze the clutter suppression performance of each optimized transmitted sequence by plotting different STAFs and cuts of different range-Doppler bins.
In Fig.\ref{shaping result} (a)-(d), we illustrate 2-D plots for the STAFs corresponding to the initial random-phased waveform and waveforms optimized by different algorithms. The target is normalized into the range-Doppler bin $(0,0)$, and the black boxes indicate the preset distribution of interfering scatters.
From these figures, we can observe that the response peaks of all STAFs are displayed at the range-Doppler bins where the target is located. Additionally, all of the designed waveforms generate deep suppression areas at the range-Doppler bins where interference points exit.
Furthermore, we notice that the response values around the range-Doppler bins of the target and interference present a certain ambiguity. This means that the formed suppression area does not strictly correspond to the predefined area of interest. This ambiguity can be explained by considering the phase error of the target in the waveform design model.
To further investigate the interference suppression performance, we employ the Doppler cuts of different STAFs. In Fig.\ref{shaping result} (c) and (c), we plot the Doppler cuts for different algorithms at $l=25, 26$. These figures demonstrate that all of the designed waveforms achieve small values for $r=11,12,\cdots,30$. Moreover, most of the response values obtained by the robust waveform are smaller than those of the other methods.

\begin{figure}
	\centering
	\subfigure[The interference distribution]{
		\begin{minipage}[t]{0.5\linewidth}
			\centering
			\includegraphics[width=4.5cm]{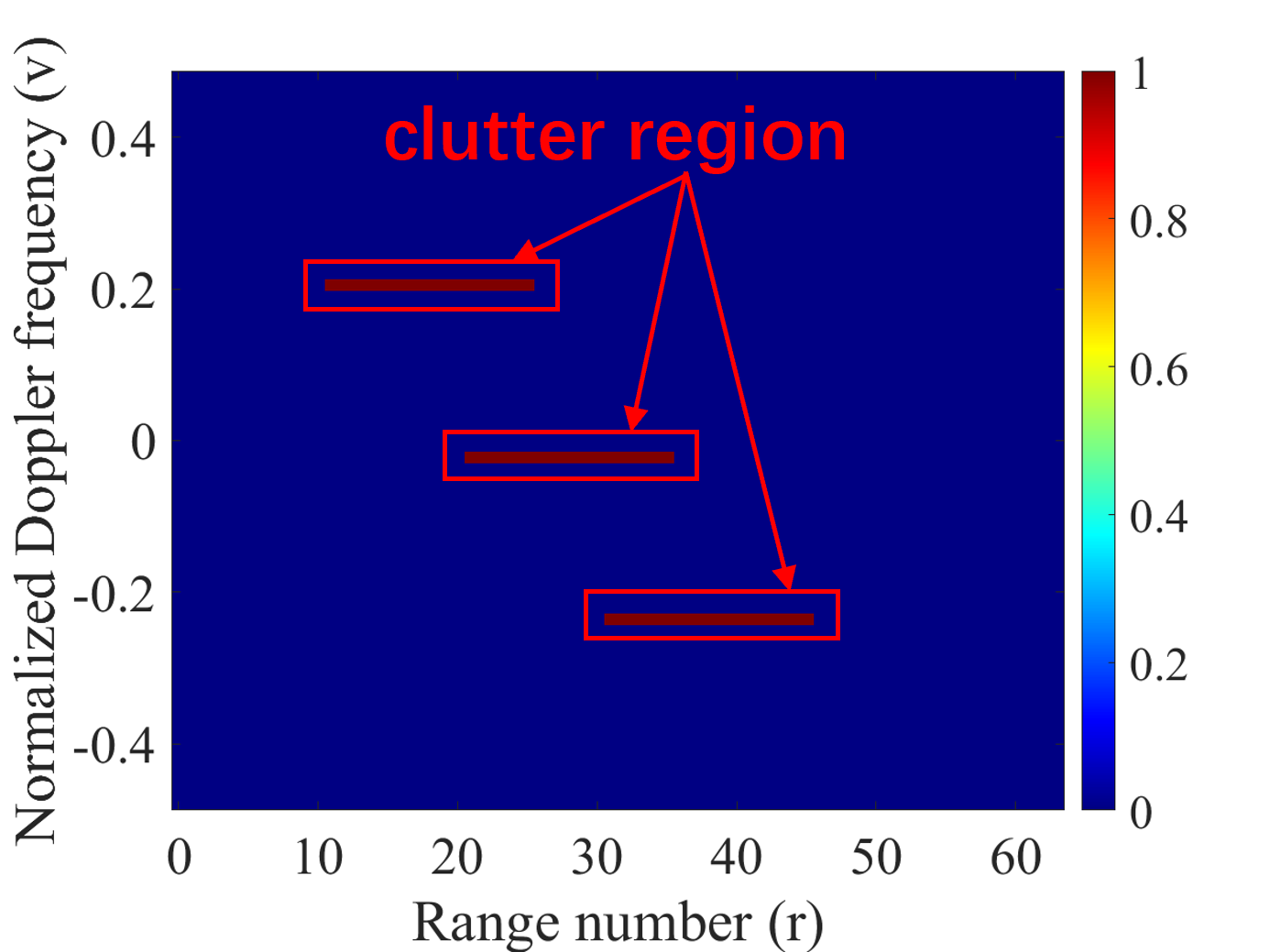}
		\end{minipage}
	}%
	\subfigure[STAF of the initial waveform]{
		\begin{minipage}[t]{0.5\linewidth}
			\centering
			\includegraphics[width=4.5cm]{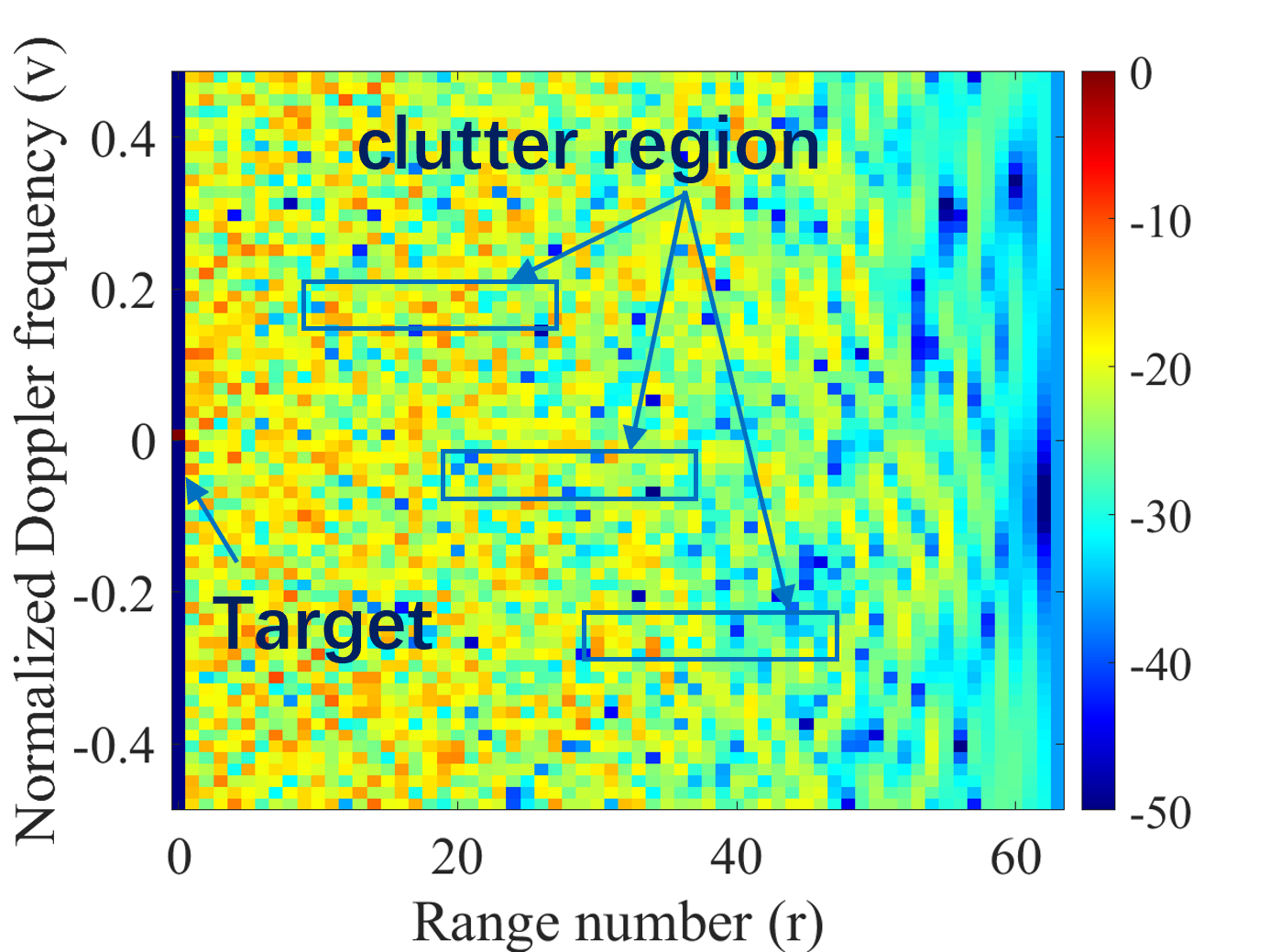}
		\end{minipage}%
	}%

 	\subfigure[STAF of the RCG waveform]{
		\begin{minipage}[t]{0.5\linewidth}
			\centering
			\includegraphics[width=4.5cm]{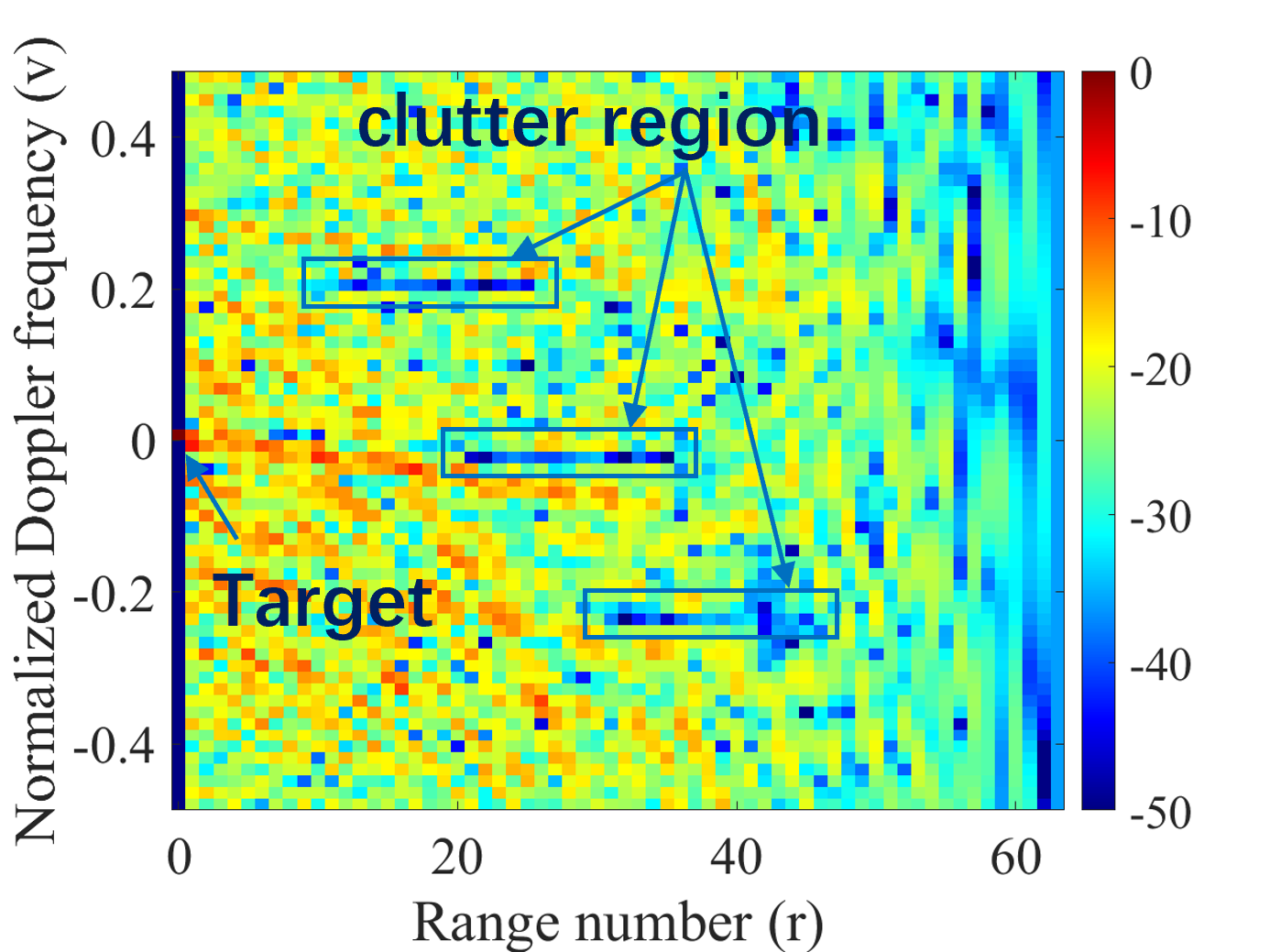}
		\end{minipage}
	}%
	\subfigure[STAF of the RTR waveform]{
		\begin{minipage}[t]{0.5\linewidth}
			\centering
			\includegraphics[width=4.5cm]{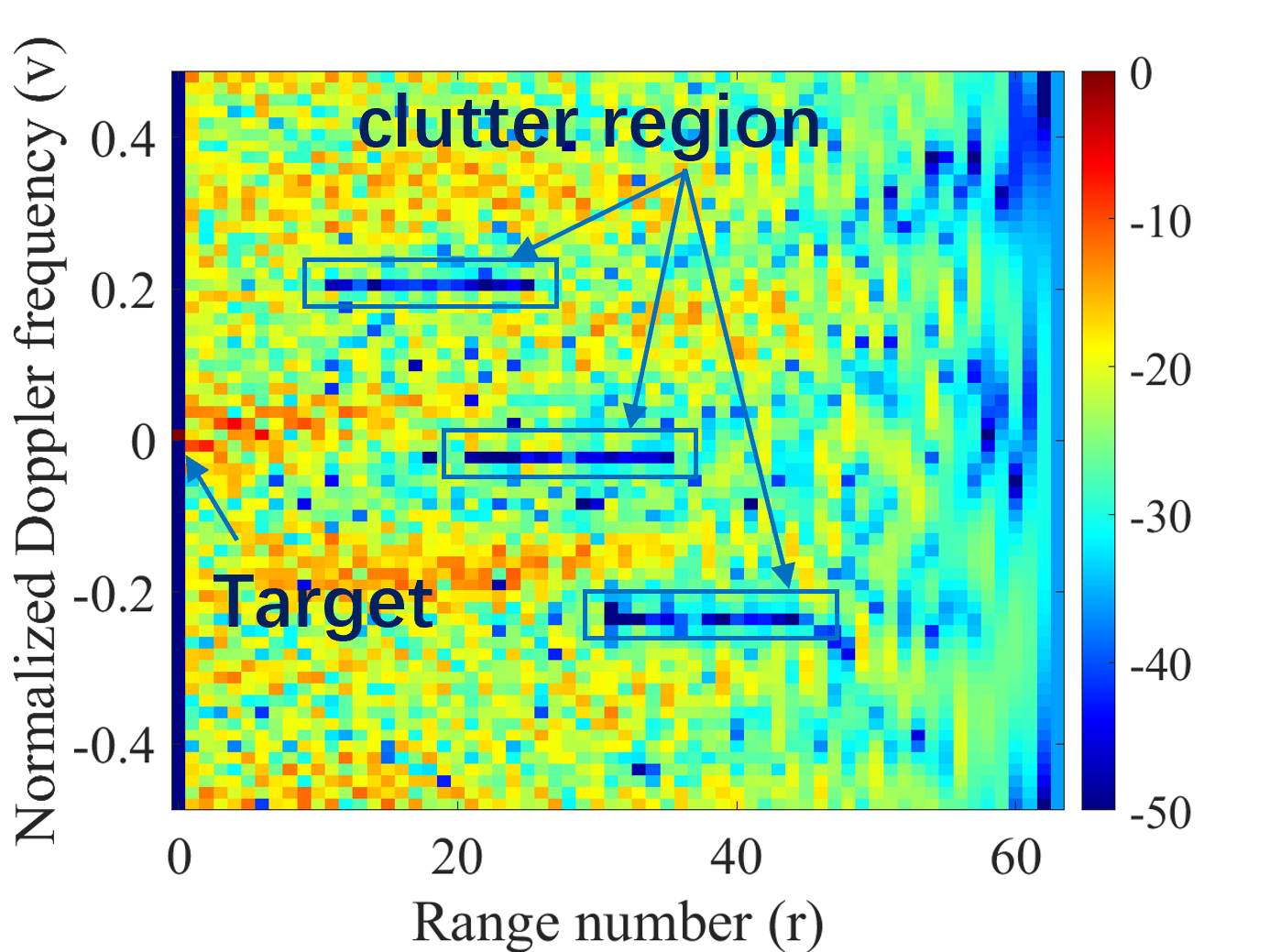}
		\end{minipage}%
	}%

  	\subfigure[STAF of the robust waveform]{
		\begin{minipage}[t]{0.5\linewidth}
			\centering
			\includegraphics[width=4.5cm]{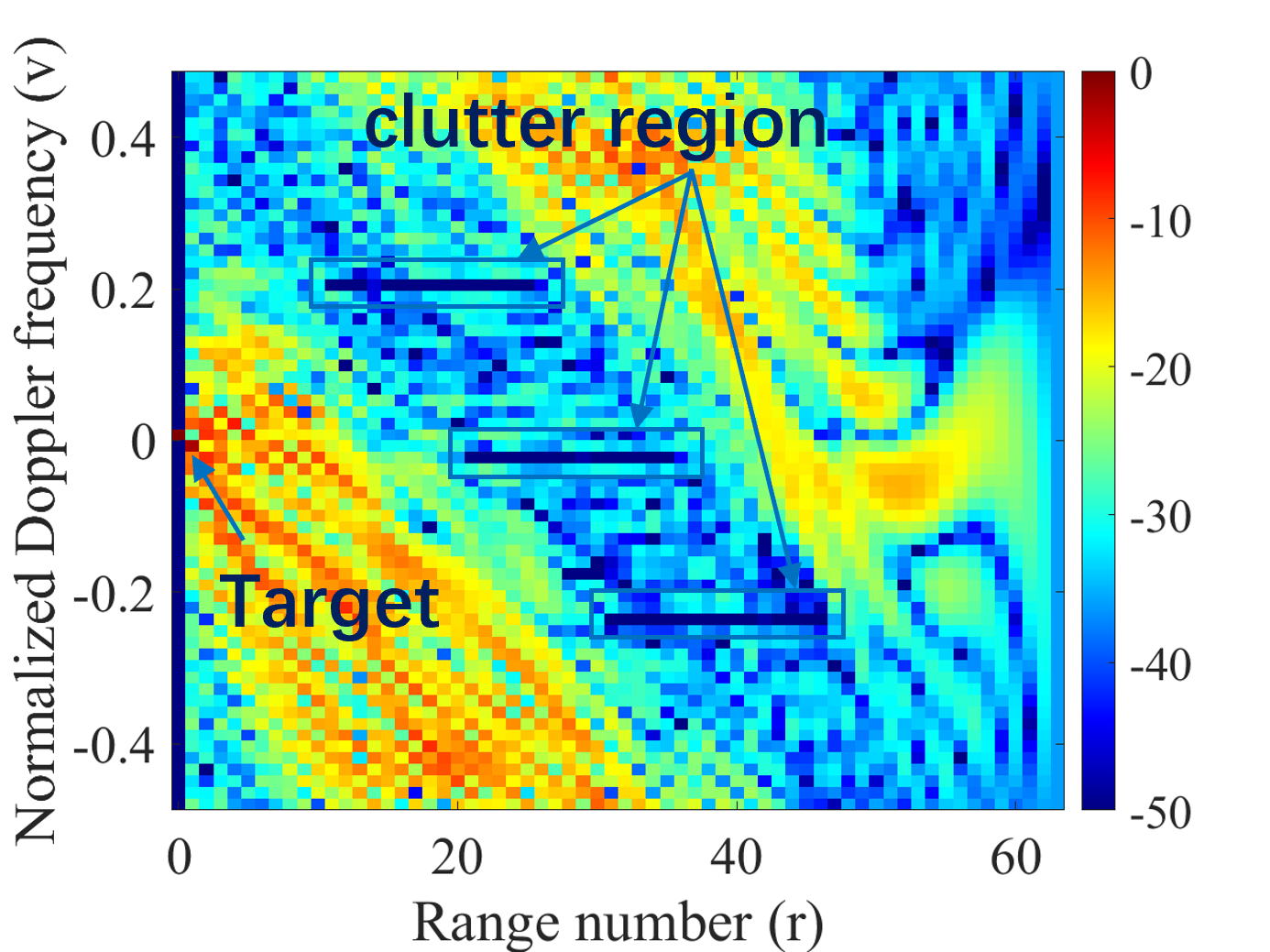}
		\end{minipage}
	}%
	\subfigure[Doppler cut of $l=16$]{
		\begin{minipage}[t]{0.5\linewidth}
			\centering
			\includegraphics[width=4.5cm]{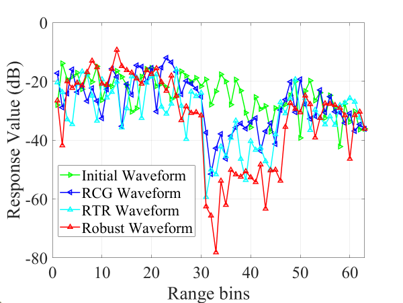}
		\end{minipage}%
	}%
 
 	\subfigure[Doppler cut of $l=30$]{
		\begin{minipage}[t]{0.5\linewidth}
			\centering
			\includegraphics[width=4.5cm]{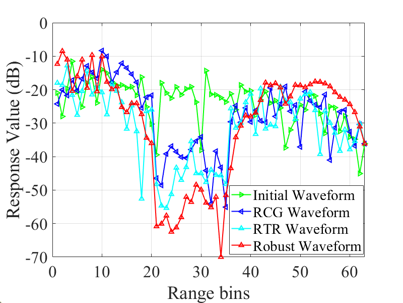}
		\end{minipage}%
	}%
  	\subfigure[Doppler cut of $l=45$]{
		\begin{minipage}[t]{0.5\linewidth}
			\centering
			\includegraphics[width=4.5cm]{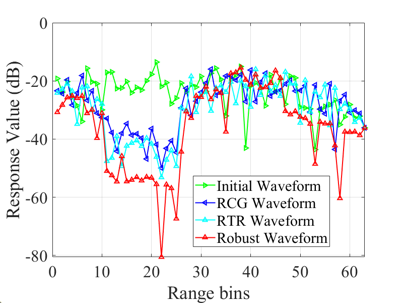}
		\end{minipage}%
	}%
	\caption{the shaping results of the second scenario.}
	\label{shaping result 2}
\end{figure}

To further demonstrate the effectiveness of our proposed method in STAF shaping, we consider a more complex scenario consisting of three subregions in the following part. Specifically, the scattering response of the interference in each range-Doppler bin can be expressed as
\begin{equation}
p(r,h)=\left\{ \begin{aligned}
& 1,\text{    }(r,h)\in \left\{16\right\}\times \left\{ 31,\cdots ,45 \right\} \\[0.1cm]
& 1,\text{    }(r,h)\in \left\{30\right\}\times \left\{ 21,\cdots ,35 \right\} \\[0.1cm] 
& 1,\text{    }(r,h)\in \left\{45\right\}\times \left\{ 11,\cdots ,25 \right\} \\[0.1cm]  
& 0,\text{    }\operatorname{otherwise} \\ 
\end{aligned}. \right.
\end{equation}
In Fig. \ref{shaping result 2} (a), we present a photographic illustration of the interference distribution. From this figure, we can observe that three interference regions with different Doppler shifts are independently distributed in different range bins.
Fig. \ref{shaping result 2} shows the STAF of the initial waveform. However, this undesigned waveform fails to form deep nulls in the interference areas, indicating unsatisfactory suppression performance.
In Figures \ref{shaping result 2} (c)-(e), we demonstrate the realized STAFs corresponding to different algorithms.
It is evident that each waveform optimization method can effectively generate deep nulls in the range-Doppler bins where the interference is located.
More precisely, the robust waveform produces a significantly deeper blue null compared to the other methods, indicating its superiority in STAF shaping.
It is important to note that the range-Doppler bins surrounding the interference areas also exhibit low response values, which can be attributed to their effectiveness in the presence of random phase error.
Additionally, we plot the range cuts in Fig. \ref{shaping result 2} (f)-(h) to compare the interference suppression performance of different algorithms. The shaped Doppler cuts of different methods exhibit deep nulls precisely where the interference is located. This further demonstrates the effectiveness of the waveform design algorithms in shaping the range-Doppler response for complex interference suppression.

\section{conclusion}
In order to optimize the slow-time transmit sequence with uncertain target steering vector, we have proposed a novel worst-case Riemannian optimization method. The proposed method transforms the original optimization problem into two sub-problems in worst-case paradigm. Each problem can be solved efficiently with the developed Riemannian algorithm whose convergence is guaranteed. Numerical simulation results show that the proposed method would converge within a few iterations and outperform the traditional methods in terms of output SCNR of the matched filter. Because the proposed method could tack with the uncertainty about the target steering vector, it is more promising in the application of cognitive radar.

\appendices
\section{Derivation of (\ref{equ12})}
\label{app1}
The manifold $\mathcal{M}$ can be expressed as
\begin{equation}
\label{ap1}
\mathcal{M} = \left\{\mathbf{x} :\mathop{Re}(\mathbf{x}(i))^{2}+\mathop{Im}(\mathbf{x}(i))^{2}=1,\forall i \in \mathcal{N}\right\}.
\end{equation}

Consider an arbitrary curve $\gamma(t)$ on the manifold. Differentiate the above equation with respect to variable $t$, we can get
\begin{equation}
\label{ap2} \mathop{Re}(\mathbf{x}(i))\mathop{Re}(\mathop{D_{t}}(\mathbf{x}(i)))+\mathop{Im}(\mathbf{x}(i))\mathop{Im}(\mathop{D_{t}}(\mathbf{x}(i)))=0.
\end{equation}
where $\mathop{D_{t}}(\cdot)$ represents the directional derivative concerning $t$ along the curve $\gamma(t)$.

Therefore the tangent vectors at $\mathbf{x}$ satisfy the following equation:
\begin{equation}
\label{ap3}
\mathop{Re}(\mathbf{\xi}_{\mathbf{x}}\odot \mathbf{x})=\mathbf{0}.
\end{equation}
where $\mathbf{\xi}_{\mathbf{x}}$ is the differential of the curve $\gamma(t)$. Since (\ref{ap1}) and (\ref{ap2}) are valid for arbitrary $\mathbf{x}$ and arbitrary curves, we can express the tangent space as (\ref{equ12}).

\section{Proof of $\textit{Theorem}\, \textit{1}$}
\label{app2}
We will prove the theorem by contradiction. First of all, let us assume $\tilde{\mathbf{s}}^{\star}$ is a plausible solution to (\ref{equ10.1}) that minimizes $\lvert\mathbf{s}^{H}\tilde{\mathbf{s}}\rvert^{2}$ and satisfies $\|\tilde{\mathbf{s}}^{\star}-\mathbf{s}\|^{2}<\varepsilon,\lvert\tilde{\mathbf{s}}^{\star}(i)\rvert=1, \forall i \in \mathcal{N}$. Let $\tilde{\mathbf{s}}^{\star} = \mathbf{s} + \Delta\mathbf{s}$. Then there exists a $\tilde{\mathbf{s}}_{1} = \mathbf{s} + \lambda\Delta\mathbf{s},\lambda \in \mathcal{R} ,\lambda>1$ that satisfies $\|\tilde{\mathbf{s}}_{1}-\mathbf{s}\|^{2}\le\varepsilon,\lvert\tilde{\mathbf{s}}_{1}(i)\rvert=1, \forall i \in \mathcal{N}$. $\tilde{\mathbf{s}}_{1}$ is also a plausible point. Then, one can get
\begin{equation}
\label{ap4}
\begin{split}
&\lvert\mathbf{s}^{H}\tilde{\mathbf{s}}^{\star}\rvert^{2}-\lvert\mathbf{s}^{H}\tilde{\mathbf{s}}_{1}\rvert^{2}\\
& = (1-\lambda^{2})\lvert\mathbf{s}^{H}\Delta\mathbf{s}\rvert^{2}+2N(1-\lambda)\mathop{Re}(\mathbf{s}^{H}\Delta\mathbf{s})
\end{split}.
\end{equation}
With the unit norm constraint, we have
\begin{equation}
\label{ap5}
\|\mathbf{s}+\Delta\mathbf{s}\|^{2}-\|\mathbf{s}+\lambda\Delta\mathbf{s}\|^{2}=0,
\end{equation}
which by derivation gives us
\begin{equation}
\label{ap6}
(1-\lambda^{2})\|\Delta\mathbf{s}\|^{2} = 2(\lambda-1)\mathop{Re}(\mathbf{s}^{H}\Delta\mathbf{s}).
\end{equation}
Substituting (\ref{ap6}) into (\ref{ap4}), one can get
\begin{equation}
\label{ap7}
\begin{split}
&\lvert\mathbf{s}^{H}\tilde{\mathbf{s}}^{\star}\rvert^{2}-\lvert\mathbf{s}^{H}\tilde{\mathbf{s}}_{1}\rvert^{2}\\
& = (1-\lambda^{2})\lvert\mathbf{s}^{H}\Delta\mathbf{s}\rvert^{2}-N(1-\lambda^{2})\|\Delta\mathbf{s}\|^{2}\\
& = (1-\lambda^{2})(\lvert\mathbf{s}^{H}\Delta\mathbf{s}\rvert^{2}-N\|\Delta\mathbf{s}\|^{2}).
\end{split}
\end{equation}
Leveraging the Cauchy-Schwarz inequality and the constraint that $\|\mathbf{s}\|^{2}=N$, we have $\lvert\mathbf{s}^{H}\Delta\mathbf{s}\rvert^{2}-N\|\Delta\mathbf{s}\|^{2}\le 0$. Therefore, it can be easily derived that $\lvert\mathbf{s}^{H}\tilde{\mathbf{s}}^{\star}\rvert^{2}-\lvert\mathbf{s}^{H}\tilde{\mathbf{s}}_{1}\rvert^{2}\ge0$. The equality holds if and only if $\Delta\mathbf{s}$ is linearly proportional to $\mathbf{s}$. Such $\tilde{\mathbf{s}}^{\star}$ clearly does not minimize the function $\lvert\mathbf{s}^{H}\tilde{\mathbf{s}}\rvert^{2}$.

Next, consider the case where the equality does not hold. Then we have $\lvert\mathbf{s}^{H}\tilde{\mathbf{s}}^{\star}\rvert^{2}-\lvert\mathbf{s}^{H}\tilde{\mathbf{s}}_{1}\rvert^{2}>0$ which contradicts the assumption that $\tilde{\mathbf{s}}^{\star}$ minimizes $\lvert\mathbf{s}^{H}\tilde{\mathbf{s}}\rvert^{2}$. Hence concludes the proof.

\section{Proof of $\textit{Theorem}\, \textit{2}$}
\label{app3}
To prove the convergence of a RTR algorithm, one only needs to check whether $\|\mathbf{grad}f_{\tilde{\mathbf{s}}}\|$, $\|\mathbf{Hess}f_{\tilde{\mathbf{s}}}\|$ and $\|\frac{D}{dt}\frac{d}{dt}\text{R}_{\tilde{\mathbf{s}}}(t\mathbf{\xi}_{\mathbf{x}})\|$ are all up bounded by certain parameters for all $\tilde{\mathbf{s}}_{k}\in \mathcal{M} $, and  $\mathbf{\xi}_{\mathbf{x}}\in T_{\mathbf{x}}\mathcal{M}$ with $\|\mathbf{\xi}_{\mathbf{x}}\|=1$, where $\frac{D}{dt}(\cdot)$ denotes the covariant derivative  along the curve $t\to R(t\mathbf{\xi}_{\mathbf{x}})$ \cite{DOI:10.1007/s10208-005-0179-9}. 

Next, we  show that $\|\mathbf{Grad}f_{\tilde{\mathbf{s}}}\|$ is up bounded, because $\|\mathbf{grad}f_{\tilde{\mathbf{s}}}\|\le\|\mathbf{Grad}f_{\tilde{\mathbf{s}}}\|$. The Euclidean gradient of algorithm 1 is presented in (\ref{equ23}). One can easily see that $\|\mathbf{Grad}f_{\tilde{\mathbf{s}}}\|$ is up bounded since we have $\|\mathbf{s}\|=1$ and $\|\tilde{\mathbf{s}}_{k}\| = 1$.

From (\ref{equ27}), we have $\|\mathbf{Hess}f_{\tilde{\mathbf{s}}}\| \le \|D{\mathbf{Grad}}f_{\tilde{\mathbf{s}}}(\mathbf{\xi}_{\mathbf{x}})\| + \|\mathop{Re}\left\{\mathbf{Grad}f_{\tilde{\mathbf{s}}}\odot\tilde{\mathbf{s}}_{k}\right\}\odot \mathbf{\xi}_{\mathbf{x}}\| $ . It is clearly up bounded,  because $\|\mathbf{\xi}_{\mathbf{x}}\|=1$.

Next, we examine $\|\frac{D}{dt}\frac{d}{dt}\text{R}_{\tilde{\mathbf{s}}}(t\mathbf{\xi}_{\mathbf{x}})\|$.  From (\ref{equ15}), we have 
\begin{equation}
\label{ap8}
\frac{d}{dt}\text{R}_{\tilde{\mathbf{s}}}(t\mathbf{\xi}_{\mathbf{x}})= \frac{\mathbf{\xi}_{\mathbf{x}}}{\|\tilde{\mathbf{s}}+t\mathbf{\xi}_{\mathbf{x}}\|}-\frac{(\tilde{\mathbf{s}}+t\mathbf{\xi}_{\mathbf{x}})\mathbf{\xi}_{\mathbf{x}}^{H}(\tilde{\mathbf{s}}+t\mathbf{\xi}_{\mathbf{x}})}{\|\tilde{\mathbf{s}}+t\mathbf{\xi}_{\mathbf{x}}\|^{3}}.
\end{equation}
It follows that  \cite{boumal2023introduction}
\begin{equation}
\label{ap9}
\begin{split}
&\frac{D}{dt}\frac{d}{dt}\text{R}_{\tilde{\mathbf{s}}}(t\mathbf{\xi}_{\mathbf{x}})\\
&= \text{Proj}_{\tilde{\mathbf{s}}_{k}}\left(\lim_{\tilde{t}\to 0}\frac{\frac{\mathbf{\xi}_{\mathbf{x}}}{\|\tilde{\mathbf{s}}+\tilde{t}\tilde{\mathbf{\xi}}_{\mathbf{x}}+t\mathbf{\xi}_{\mathbf{x}}\|}-\frac{\mathbf{\xi}_{\mathbf{x}}}{\|\tilde{\mathbf{s}}+t\mathbf{\xi}_{\mathbf{x}}\|}}{dt}\right)\\
&-\text{Proj}_{\tilde{\mathbf{s}}_{k}}\left(\lim_{\tilde{t}\to 0}\frac{\frac{(\tilde{\mathbf{s}}+\tilde{t}\tilde{\mathbf{\xi}}_{\mathbf{x}}+t\mathbf{\xi}_{\mathbf{x}})\mathbf{\xi}_{\mathbf{x}}^{H}(\tilde{\mathbf{s}}+\tilde{t}\tilde{\mathbf{\xi}}_{\mathbf{x}}+t\mathbf{\xi}_{\mathbf{x}})}{\|\tilde{\mathbf{s}}+\tilde{t}\tilde{\mathbf{\xi}}_{\mathbf{x}}+t\mathbf{\xi}_{\mathbf{x}}\|^{3}}-\frac{(\tilde{\mathbf{s}}+t\mathbf{\xi}_{\mathbf{x}})\mathbf{\xi}_{\mathbf{x}}^{H}(\tilde{\mathbf{s}}+t\mathbf{\xi}_{\mathbf{x}})}{\|\tilde{\mathbf{s}}+t\mathbf{\xi}_{\mathbf{x}}\|^{3}}}{dt}\right)\\
&= \text{Proj}_{\tilde{\mathbf{s}}_{k}}\left(\frac{(\tilde{\mathbf{s}}+t\mathbf{\xi}_{\mathbf{x}})\mathbf{\xi}_{\mathbf{x}}^{H}\tilde{\mathbf{\xi}}_{\mathbf{x}}}{\|\tilde{\mathbf{s}}+t\mathbf{\xi}_{\mathbf{x}}\|^{3}}-\frac{3\|\tilde{\mathbf{s}}+t\mathbf{\xi}_{\mathbf{x}}\|\tilde{\mathbf{\xi}}_{\mathbf{x}}^{H}(\tilde{\mathbf{s}}+t\mathbf{\xi}_{\mathbf{x}})}{\|\tilde{\mathbf{s}}+t\mathbf{\xi}_{\mathbf{x}}\|^{6}}\right).\\
\end{split}
\end{equation}
Considering that both $\|\tilde{\mathbf{\xi}}_{\mathbf{x}}\|=1$ and $\|\mathbf{\xi}_{\mathbf{x}}\|=1$, we find that  $\frac{D}{dt}\frac{d}{dt}\text{R}_{\tilde{\mathbf{s}}}(t\mathbf{\xi}_{\mathbf{x}})$ is clearly up bounded when $t<\delta_{t}$. Therefore, with the proposition given in  \cite{DOI:10.1007/s10208-005-0179-9}, it holds that
\begin{equation}
\label{ap10}
\lim_{k\to \infty}\mathbf{grad}f_{\tilde{\mathbf{s}}}(\tilde{\mathbf{s}}_{k})=\mathbf{0}
\end{equation}

\section{Proof of $\textit{Theorem}\, \textit{3}$}
\label{app4}
Let $\mathbf{v}$ be a nondegenerate local minimizer of $f_{\mathbf{s}}$, i.e., $\mathbf{grad}f_{\mathbf{s}}(\mathbf{v})=\mathbf{0}$ and $\mathbf{Hess}f_\mathbf{s}(\mathbf{v})$ is positive definite. Now, we will prove that there exists a neighborhood $\mathbf{V}$ of $\mathbf{v}$ such that, for all $\mathbf{s}_{0}\in \mathbf{V}$, the sequence $\{\mathbf{s}_{k}\}$ generated converges to $\mathbf{v}$. To prove the above theorem, we need to check the following conditions: i) $f_{\mathbf{s}}$ is $\mathcal{C}^{1}$ and bounded below on the level set $\{\mathbf{s}\in \mathcal{M}:f_{\mathbf{s}}(\mathbf{s})\le f_{\mathbf{s}}(\mathbf{s}_{0})\}$ where $\mathbf{s}_{0}$ is the initial point; ii) $f_{\mathbf{s}}(\textbf{R}_{\mathbf{x}}(\tau \mathbf{\xi}_{\mathbf{x}}))$ is radially $\mathcal{L}-\mathcal{C}^{1}$, i.e. radially Lipschitz continuously differentiable; iii) $\|\mathbf{Hess}f_{\mathbf{s}}\|_{\mathcal{o}}\le \beta$ for some constant $\beta$ and $\mathbf{s}\to \|\mathbf{Hess}f_{\mathbf{s}}^{-1}\|_\mathcal{o}$ is bounded on a neighborhood of $\mathbf{v}$, where $\|\mathbf{Hess}f_{\mathbf{s}}\|_{\mathcal{o}}:= \text{sup}\left\{\|\mathbf{Hess}f_{\mathbf{s}}(\mathbf{\xi})\|: \mathbf{\xi} \in T_{\mathbf{x}}\mathcal{M},\|\mathbf{\xi}\| = 1\right\}$; iv) there exists some $\mu>0$ and $\delta_{\mu}>0$ such that $\|\mathbf{\xi}\|\ge \mu \text{dist}(\mathbf{x}, \textbf{R}_{\mathbf{x}}(\mathbf{\xi}))$ for all $\mathbf{x}\in \mathcal{M}$ and for all $\mathbf{\xi}\in T_{\mathbf{x}}\mathcal{M}$, $\|\mathbf{\xi}\|\le \delta_{\mu}$ where $\text{dist}(\mathbf{x}, \textbf{R}_{\mathbf{x}}(\mathbf{\xi}))$ is the Riemannian distance between $\mathbf{x}$ and $\textbf{R}_{\mathbf{x}}(\mathbf{\xi})$  \cite{DOI:10.1007/s10208-005-0179-9}. 

The first condition is obviously satisfied and skipped to the next. To prove $f_{\mathbf{s}}$ is radially $\mathcal{L}-\mathcal{C}^{1}$, we need to compute $\frac{df_{\mathbf{s}}(\textbf{R}(\tau\mathbf{\xi}))}{d\tau}$ with $\mathbf{\xi} \in T_{\mathbf{x}}\mathcal{M} $ and $\|\mathbf{\xi}\| = 1$:
\begin{equation}
\label{ap11}
\begin{split}
\frac{df_{\mathbf{s}}(\textbf{R}(\tau\mathbf{\xi}))}{d\tau} & =\lim_{\tau \to 0} \frac{\frac{\sum_{i=0}^{N_{t}-1}\lvert(\mathbf{s}^{H}+\tau\mathbf{\xi})\mathbf{\Psi}_{i}(\mathbf{s}+\tau\mathbf{\xi})\rvert^{2}}{\lvert(\mathbf{s}+\tau\mathbf{\xi})^{H}\tilde{\mathbf{s}}\rvert^{2}\|\mathbf{s}+\tau\mathbf{\xi}\|^{2}}-\frac{\sum_{i=0}^{N_{t}-1}\lvert\mathbf{s}^{H}\mathbf{\Psi}_{i}\mathbf{s}\rvert^{2}}{\lvert\mathbf{s}^{H}\tilde{\mathbf{s}}\rvert^{2}}}{\tau}\\
& = \frac{2\mathbf{s}^{H}\mathbf{\Psi}_{i}^{H}\mathbf{s}(\mathbf{\xi}^{H}\mathbf{\Psi}_{i}\mathbf{s}+\mathbf{s}^{H}\mathbf{\Psi}_{i}\mathbf{\xi})}{\lvert\mathbf{s}^{H}\tilde{\mathbf{s}}\rvert^{2}}\\
&\, - \frac{2\vert\mathbf{s}^{H}\mathbf{\Psi}_{i}\mathbf{s}\rvert^{2}\mathbf{s}^{H}\mathbf{\xi}}{\lvert\mathbf{s}^{H}\tilde{\mathbf{s}}\rvert^{2}}\\
&\,- \frac{2\vert\mathbf{s}^{H}\mathbf{\Psi}_{i}\mathbf{s}\rvert^{2}\mathbf{s}^{H}\tilde{\mathbf{s}}\tilde{\mathbf{s}}^{H}\mathbf{\xi}}{\lvert\mathbf{s}^{H}\tilde{\mathbf{s}}\rvert^{4}}.
\end{split}
\end{equation}
Note that we have assumed that our initial point starts within a neighborhood of the local minimizer $\mathbf{v}$. Therefore, there exists a submanifold of $\mathcal{M}$ defined by such neighborhood that meets $\lvert\mathbf{s}^{H}\tilde{\mathbf{s}}\rvert\ge\delta_{\tilde{\mathbf{s}}}$ for some constant $\delta_{\tilde{\mathbf{s}}}$. With this assumption, one could easily found that $\frac{df_{\mathbf{s}}(\textbf{R}(\tau\mathbf{\xi}))}{d\tau}$ is continuous and bounded. Following the same paradigm as in (\ref{ap11}), one could check that its gradient at $\tau = 0$ exists and up bounded, i.e. 
\begin{equation}
\label{ap12}
\bigg|\lim_{t\to 0}\frac{\frac{df_{\mathbf{s}}(\textbf{R}(\tau\mathbf{\xi}))}{d\tau}\lvert_{\tau=t}-\frac{df_{\mathbf{s}}(\textbf{R}(\tau\mathbf{\xi}))}{d\tau}\lvert_{\tau=0}}{t}\bigg| \le \delta_{\tau}.
\end{equation}
Thus, the function $f_{\mathbf{s}}(\textbf{R}_{\mathbf{x}}(\tau \mathbf{\xi}_{\mathbf{x}}))$ is radially $\mathcal{L}-\mathcal{C}^{1}$, i.e. 
\begin{equation}
\label{ap13}
\bigg|\frac{df_{\mathbf{s}}(\textbf{R}(\tau\mathbf{\xi}))}{d\tau}\lvert_{\tau=t}-\frac{df_{\mathbf{s}}(\textbf{R}(\tau\mathbf{\xi}))}{d\tau}\lvert_{\tau=0}\bigg|\le \beta_{RL}t
\end{equation}
for all $\mathbf{s}$ that meets $\lvert\mathbf{s}^{H}\tilde{\mathbf{s}}\rvert\ge\delta_{\tilde{\mathbf{s}}}$ and for all $\mathbf{\xi}\in T_{\mathbf{x}}\mathcal{M}$ with $\|\mathbf{\xi}\|=1$ and $t<\delta_{RL}$.  Hence, condition ii) is satisfied. 
From (\ref{equ39}), it can be found that $\|\mathbf{Hess}f_{\mathbf{s}}\|_{\mathcal{o}}$ is up bounded when $\lvert\mathbf{s}^{H}\tilde{\mathbf{s}}\rvert\ge\delta_{\tilde{\mathbf{s}}}$. Therefore, when we only consider a submanifold that is composed of a neighborhood near the local minimizer, we have $\|\mathbf{Hess}f_{\mathbf{s}}\|_{\mathcal{o}}\le\beta$. Since $\mathbf{Hess}f_\mathbf{s}(\mathbf{v})$ is positive definite, the smallest eigenvalue of the operator $\mathbf{Hess}f_\mathbf{s}$ is positive.  Thus it is trivial to prove that $\mathbf{s}\to \|\mathbf{Hess}f_{\mathbf{s}}^{-1}\|_\mathcal{o}$ is bounded on a neighborhood of $\mathbf{v}$.  Therefore, condition iii) is satisfied.
Condition iv) states that the length of a tangent vector is always proportionally larger than the Riemannian distance , i.e.  the length of the geodesic between two points. For the complex circle manifold discussed in this paper, one could easily check this condition is hold, because the geodesic is our case is the arc between two points. This bound is known to be satisfied for the exponential retraction discussed in this paper  \cite{DOI:10.1007/s10208-005-0179-9}. Also this bound is satisfied when the retraction is smooth and $\mathcal{M}$ is compact. 
Now, we have proved that all the above conditions hold. Therefore, we have that for all $\mathbf{s}_{0}\in \mathbf{V}$, the sequence generated by algorithm 2 converges to $\mathbf{v}$. Hence concludes the proof.




\bibliographystyle{IEEEtran}
\bibliography{ref}

\begin{IEEEbiography}
	[{\includegraphics[width=1in,height=1.25in, clip,keepaspectratio]{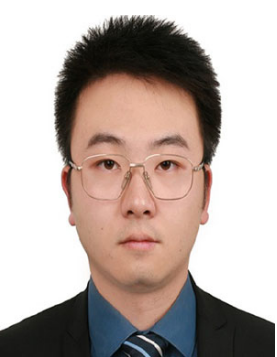}}]{Xinyu Zhang} received the B.S. and Ph.D. degrees from the Beijing Institute of Technology, Beijing, China, in 2011 and 2017, respectively. He is currently a Lecturer at the National University of Defense Technology. He visited the Ohio State University from 2015 to 2017 as a Visiting Scholar. Since 2017, he has joined the National University of Defense Technology as a Post-doctor. His research interests include array signal processing, auto-target detection, and waveform optimization.
\end{IEEEbiography}

\begin{IEEEbiography}[{\includegraphics[width=1in,height=1.25in,clip,keepaspectratio]{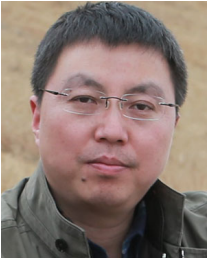}}]{Weidong Jiang}	was born in Chongqing, China, in 1968. He received a B.S. degree in communication engineering and a Ph.D. degree in electronic science and technology from the National University of Defense Technology (NUDT), China, in 1991 and 2001, respectively. He is currently a Professor with NUDT. His current research interests include MIMO radar signal processing and radar system technology.
\end{IEEEbiography}

\begin{IEEEbiography}[{\includegraphics[width=1in,height=1.25in,clip,keepaspectratio]{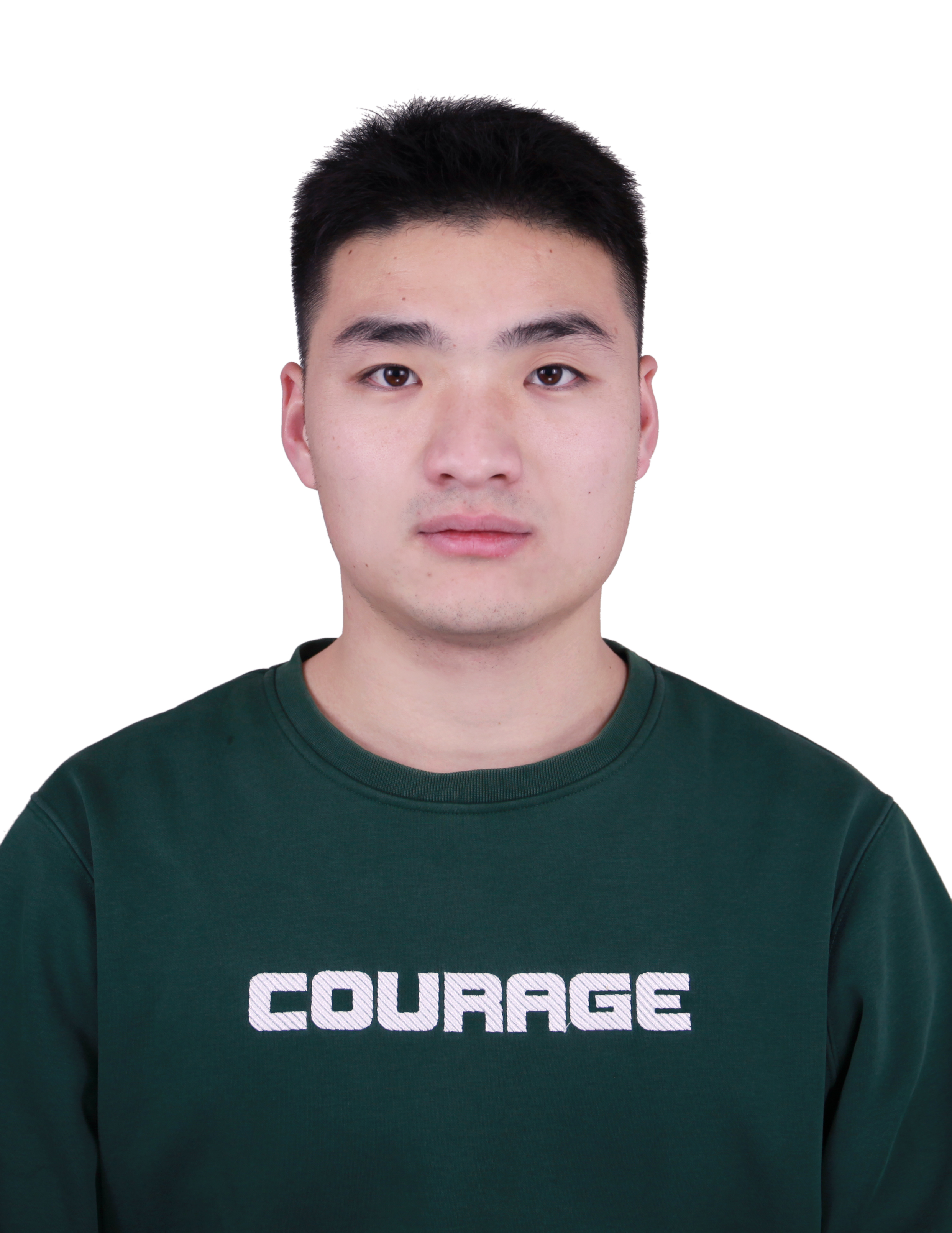}}]{Xiangfeng Qiu}	was born in Anhui, China, in 1997. He received a B.S. degree in Information and Communication Engineering from the College of Electronic Science and Technology at the National University of Defense Technology (NUDT), China, in 2021. His current research interests include radar waveform design and radar target recognition.
\end{IEEEbiography}

\begin{IEEEbiography}[{\includegraphics[width=1in,height=1.2in,clip,keepaspectratio]{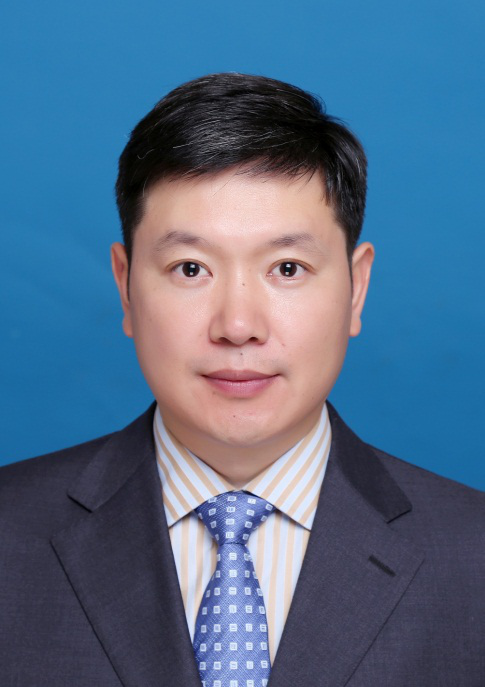}}]{Yongxiang Liu} received the B.S. and Ph.D. degrees from the National University of Defense Technology (NUDT), China, in 1999 and 2004, respectively. Since 2004, he has been with NUDT and is currently a Professor with the College of Electrical Science and Engineering, conducting research on radar target recognition, time-frequency analysis, micro motions, and array signal processing. He worked at Imperial College London as an Academic Visitor in 2008.
\end{IEEEbiography}
\end{document}